\documentclass[preprint2,times,tighten]{aastex6}
\usepackage{amsmath,amssymb,amstext}
\usepackage[figure,figure*]{hypcap}
\usepackage{newtxmath} 

\graphicspath{{./}{figs/}}

\newcommand{\angstrom}{{\rm \mathring A}}

\shorttitle{EUCLIA - Exploring the UV/optical continuum lag in active galactic nuclei}
\shortauthors{Cai et al.}

\begin{document}
\title{
EUCLIA - Exploring the UV/optical continuum lag in active galactic nuclei. I. a model without light echoing
}
\author{Zhen-Yi Cai\altaffilmark{1,2}, Jun-Xian Wang\altaffilmark{1,2}, Fei-Fan Zhu\altaffilmark{1,2,3}, Mou-Yuan Sun\altaffilmark{1,2}, Wei-Min Gu\altaffilmark{4,7}, Xin-Wu Cao\altaffilmark{5,6,7}, Feng Yuan\altaffilmark{5,7}}
\altaffiltext{1}{CAS Key Laboratory for Research in Galaxies and Cosmology, Department of Astronomy, University of Science and Technology of China, Hefei 230026, China; zcai@ustc.edu.cn, jxw@ustc.edu.cn}
\altaffiltext{2}{School of Astronomy and Space Science, University of Science and Technology of China, Hefei 230026, China}
\altaffiltext{3}{Department of Astronomy and Astrophysics, 537 Davey Lab, The Pennsylvania State University, University Park, PA 16802, USA}
\altaffiltext{4}{Department of Astronomy, Xiamen University, Xiamen, Fujian 361005, China}
\altaffiltext{5}{Shanghai Astronomical Observatory, Chinese Academy of Sciences, 80 Nandan Road, Shanghai 200030, China}
\altaffiltext{6}{Key Laboratory of Radio Astronomy, Chinese Academy of Sciences, Nanjing 210008, China}
\altaffiltext{7}{SHAO-XMU Joint Center for Astrophysics, Xiamen University, Xiamen, Fujian 361005, China}

\begin{abstract}
The tight inter-band correlation and the lag-wavelength relation among UV/optical continua of active galactic nuclei have been firmly established. They are usually understood within the widespread reprocessing scenario, however, the implied inter-band lags are generally too small.
Furthermore, it is challenged by new evidences, such as the X-ray reprocessing yields too much high frequency UV/optical variations as well as it fails to reproduce the observed timescale-dependent color variations among {\it Swift} lightcurves of NGC 5548.
In a different manner, we demonstrate that an upgraded inhomogeneous accretion disk model, whose local {\it independent} temperature fluctuations are subject to a speculated {\it common} large-scale temperature fluctuation, can intrinsically generate the tight inter-band correlation and lag across UV/optical, and be in nice agreement with several observational properties of NGC 5548, including the timescale-dependent color variation. 
The emergent lag is a result of the {\it differential regression capability} of local temperature fluctuations when responding to the large-scale fluctuation. 
An average speed of propagations as large as $\gtrsim 15\%$ of the speed of light may be required by this common fluctuation. Several potential physical mechanisms for such propagations are discussed.
Our interesting phenomenological scenario may shed new light on comprehending the UV/optical continuum variations of active galactic nuclei.
\end{abstract}


\section{Introduction}

More and more efforts have been devoted to understand the ubiquity of the variability of active galactic nuclei (AGNs) from radio waves to X-rays and gamma-rays \citep[e.g.,][]{Wallinder1992,Ulrich1997,UttleyCasella2014,Xue2017}, which in turn are improving our comprehension of the properties of the structure of AGNs, such as the sizes of the accretion disk as well as the broad line region, the temperature profile of the disk, the physical processes over the disk, the interactions between the inner hot corona and the outer cold disk, and so on \citep[e.g.,][]{Fabian2009,GardnerDone2014,Sun2015,Buisson2017}. There are many observational properties of the variability, including the relationships between the UV/optical variations and the physical properties of AGNs (e.g., the wavelength, the bolometric luminosity, the Eddington ratio, the black mass; \citealt{Macleod2010,Guo2017}, and the introduction of \citealt{Cai2016} for more detailed references), the so-called (timescale-dependent) bluer-when-brighter trend \citep[e.g.,][]{Schmidt2012,Sun2014,Guo2016}, and the coordinations across the UV/optical continua \citep[e.g.,][and references therein]{Lawrence2012,Shappee2014,Fausnaugh2016}.


Our current interests primarily focus on the subject of the coordination property among the UV/optical continua, that is, the continuum variation at longer wavelength is not only tightly correlated with but also lags that at shorter wavelength, and the lag increases with increasing wavelength. 
The tight inter-band correlations among continua have been claimed as early as about 1990s in a handful of local Seyfert galaxies \citep[e.g.,][]{Clavel1991,Krolik1991,Korista1995,Crenshaw1996,Edelson1996,Peterson1998,Edelson2000,Sergeev2005} and several hundred more luminous quasars \citep[e.g.,][]{Giveon1999,Hawkins2003}, while the positive lag-wavelength relation was firstly significantly detected in the Seyfert 1 galaxy NGC 7469 by \citet{Wanders1997}, later confirmed and extended to longer wavelengths by \citet{Collier1998} and \citet{Kriss2000}. 
Later on, evidences for the lag-wavelength relation were gradually reported by many other studies \citep[e.g.,][]{Collier2001,Doroshenko2005,Sergeev2005}. 
Particularly, thanks to the modern unprecedented monitoring campaigns of a few AGNs from X-ray to near-infrared, including NGC 2617 \citep{Shappee2014}, NGC 5548 \citep{McHardy2014,DeRosa2015,Edelson2015,Fausnaugh2016}, NGC 4395 \citep{McHardy2016}, NGC 4151 \citep{Edelson2017}, and NGC 4593 \citep{Cackett2017,McHardy2017}, this lag-wavelength relation has been firmly established, though still in a few sources. 


Long before its convincing discovery, this lag-wavelength relation has been expected within the famous hard X-ray reprocessing models \citep{Krolik1991}. These models envisage an outer cold accretion disk \citep[e.g.,][]{ShakuraSunyaev1973} whose surface layer reprocesses, i.e., heated by and quickly re-radiating thermally \citep[e.g.,][]{Starkey2017}, the incident variable hard X-ray illumination from an inner hot corona \citep{HaardtMaraschi1991,ReynoldsNowak2003}. 
With increasing the disk radius, the response of disk to the hard X-ray results in an increasing time delay at longer wavelengths, comparable to the light-travel time from the central corona. 
Since its speculation, this reprocessing scenario has been widely assumed and reformed to make quantitative comparisons to observations \citep[e.g.,][]{Edelson1996,Collier2001,KoristaGoad2001,Frank2002book,Sergeev2005,Cackett2007,Breedt2009,McHardy2014,Uttley2014,Fausnaugh2016,Noda2016,Starkey2017,GardnerDone2017}.


The reprocessing scenario is potentially successful on several aspects, besides its simple physical origin, while it is also challenged as old as its recommendation and by new evidences enriched recently.
Firstly, the overall correlation between the X-ray and UV/optical bands as well as the variations in UV/optical bands lagging those in X-ray bands by only one to several days are found in several galaxies \citep[e.g.,][]{McHardy2014,Edelson2015,Troyer2016}, but not confirmed in a few different galaxies observed for days to years \citep[e.g.,][]{Edelson2000,Arevalo2005,SmithVaughan2007,Lira2015}.
Secondly, the connection between X-ray and UV/optical is also planted by the similar shapes of power spectral density found across X-ray to UV/optical on timescales of $\sim 5-60$ days in about ten AGNs \citep{CollierPeterson2001}, together with the positive correlation between X-ray and UV luminosities derived from several hundred of AGNs \citep[e.g.,][]{Vignali2003,Strateva2005,Steffen2006,LussoRisaliti2016}. 
However, more debates come from the delicate comparisons between the hard X-ray and UV/optical light curves. 
On one hand, the UV/optical light curves are not well reproduced, especially at timescales of hundreds of days, when assuming that the driving light curve for the UV/optical variations is the observed hard X-ray light curve \citep[e.g.,][]{KazanasNayakshin2001,Breedt2009,Shappee2014}.
On the other hand, the plausible driving X-ray light curve directly inferred from the UV/optical light curves is found to correlate poorly with both the observed hard and soft X-ray light curves \citep[e.g.,][]{Starkey2016,Starkey2017}.
In other words, too much short-term variabilities inheriting from the X-ray are preserved in the UV/optical light curves as well as the smaller lag times are lingering within the reprocessing scenario \citep{GardnerDone2017}.
Thirdly, although the reprocessing scenario naturally implies a lag-wavelength dependence of $\tau \propto \lambda^{4/3}$ for a standard thin disk from UV to IR bands, the predicted UV/optical lags are generally smaller than those of observations \citep[e.g.,][]{Shappee2014,Fausnaugh2016}, unless referring to a much higher accretion rate, a larger disk size at given wavelength, a steeper profile of disk temperature than those of the standard thin disk, or/and a change of disk structure when it is under illumination \citep[e.g.,][]{McHardy2014,Lira2015,Troyer2016,Fausnaugh2016,Starkey2017,GardnerDone2017}.
Last but not least, new evidence against the reprocessing scenario is its failure in reproducing the observed timescale-dependent color variations among {\it Swift} lightcurves of NGC 5548 \citep{Zhu2017}.
Another point to be mentioned is the potential tension between the monotonic increasing lag with wavelength implied by the reprocessing and the possible flattening of the observed lags at the longest wavelengths \citep{Fausnaugh2016}.
Consequently, the UV/optical variations may not be driven by the hard X-ray illumination, especially when the X-ray/UV luminosity ratio is small and indeed this ratio decreases with increasing UV luminosity for luminous quasars \citep[e.g.,][]{Strateva2005}.
All these challenges may suggest that there exists some other dominant mechanisms, rather than reprocessing of hard X-ray, responsible for the UV/optical variability \citep[cf.][]{Edelson2000,Breedt2009,McHardy2014,Starkey2017}. 
All in all, the physical origin of the UV/optical continuum variability remains unclear.

Indeed, to overcome the dilemma of too much fast variability encountered by the hard X-ray reprocessing models, \citet{GardnerDone2017} propose the solution could be the reprocessing of the far-UV emissions, the emitting region of which is a relatively outer part of an inflated inner accretion disk.
The observed soft X-ray excess is then attributed to the relatively inner part of this inflated inner disk, which itself shields the most inner region of hard X-ray emission from illuminating the outer thin disk. 
However, the resulting lags from the thin disk contributing to the UV/optical emissions are still too short. As they suggest, the extra lags may originate from the differential responsibility of the disk vertical structure to the changing of far-UV illuminating.
Note that a similar scenario is also speculated by \citet{Lawrence2012} where the inner disk with extreme-UV emission heats the outer disk as well as the a population of cold, thick clouds in order to solve the problems appertained to the UV bump in AGNs, including the temperature, ionization, timescale, and coordination problems (cf. the Section 2 of \citealt{Lawrence2012} for more details).


Nevertheless, all the above efforts are faithfully relying on the reprocessing scene though somewhat different. Motivated by several other suggestions from observations that the UV/optical variations may have different characteristic timescales, potentially driven by completely different mechanisms \citep[e.g.,][]{Edelson2000,Breedt2009,McHardy2014}, we start to seek whether there is a new solution without any reprocessing. 
The most easily revivable picture is the temperature fluctuations on accretion disk, as a result of some disk instabilities \citep[e.g.,][]{Kawaguchi1998}. 
Considering the inhomogeneous disk model, where local temperatures highly fluctuate, \citet{DexterAgol2011} initially propose to simultaneously explain the observed larger size of the accretion disk and the amplitudes of the optical variability, and \citet{Ruan2014} further demonstrate its better responsibility for the bluer-when-brighter trend constructed from a sample of variable quasars.
In \citet{Cai2016}, we also reveal its compatibility with the observed timescale-dependent color variations, that short-term variations are even bluer than longer term ones at all redshifts up to $z \sim 3.5$, firstly discovered by \citet{Sun2014} using SDSS $g/r$ band light curves and confirmed by \citet{Zhu2016} adopting {\it GALEX} far- and near-UV light curves for a sample of quasars. Furthermore, this timescale-dependent color variation is amazingly detected for the first time in an individual AGN, i.e., NGC 5548, with high quality {\it Swift} multi-band UV/optical light curves \citep{Zhu2017}.


Being disappointed, as pointed out by previous studies \citep[e.g.,][]{Kokubo2015,Cai2016}, is the facts that both the inter-band correlations become weaker correlated with increasing the wavelength difference of two bands and, more seriously, completely no lags among the UV/optical wavelengths are implied within the earlier versions of the inhomogeneous models, where the fluctuations at different disk regions are assumed to be totally independent to each other.
Obviously, these earlier models are oversimplified and in reality the connection among different disk regions is plausible at least due to the propagation of fluctuations induced by instabilities across the disk. There may be many distinct outward/inward propagations at different timescales and then a net effect could be that a smoothed common large-scale fluctuation emerges across a large disk region beyond a characteristic timescale related to an average speed of propagations. 
Up to now, the propagations have primarily been understood on theoretical point of view, but fortunately observations on the so-called changing-look AGNs, experiencing dramatic changes of X-ray/UV/optical flux and spectral class
on a timescale of $\sim 1-10$ years, are now stimulating our understanding on this direction \citep[e.g.,][]{Shappee2014,Denney2014,LaMassa2015,McElroy2016,Runco2016,Runnoe2016,MacLeod2016,Ruan2016,Gezari2017,Sheng2017}.
The changes of flux and state are generally attributed to the change of accretion rate. Nevertheless, the fact that for example the source reported by \citet{Gezari2017} the observed changing timescale is probably smaller by about three order of magnitude than the viscous radial drift timescale may indicate that some quick propagations across the disk may have happened in order to generate a large-amplitude outburst on a short timescale.

Since the difficulty in explaining the tight correlations and the wavelength-dependent time lag across the UV/optical wavelengths may be attributed to the too strong assumption that temperature fluctuations in different disk zones are completely independent, we strive to consider such a speculated common large-scale fluctuation over the whole disk as an exploratory model, mimicking the possible real propagation of fluctuations, outlined in Section~\ref{sect:model}. 
Encouragingly, we are able to recover a nice agreement with several observations of NGC 5548, including the tight correlations and comparable lags as illustrated in Section~\ref{sect:results}.
Followed are discussions on several properties of the light curves and the potential physical mechanisms for our speculated common large-scale fluctuation in Sections~\ref{sect:discussion} and \ref{sect:phy_mech}, respectively.
Finally, a brief conclusion is presented in Section~\ref{sect:conclusion}.

Throughout the paper we will mainly compare the model to the new data of NGC 5548 \citep{DeRosa2015,Edelson2015,Fausnaugh2016} with redshift $z = 0.017175$ (or luminosity distance $D_{\rm L} = 77.775$ Mpc), black hole mass $M_{\bullet} = 5 \times 10^7 M_\odot$ \citep{Bentz2007,Woo2010,Grier2013b,Pancoast2014}, and Eddington ratio $\lambda_{\rm Edd} = 0.1$ as reference.
The comparisons with the other sources will be deferred in this series of exploring the UV/optical continuum lag in AGN.

\section{Inhomogeneous accretion disk model with a speculated common large-scale fluctuation}\label{sect:model}

\begin{figure*}[tb!]
\centering
\includegraphics[width=\textwidth]{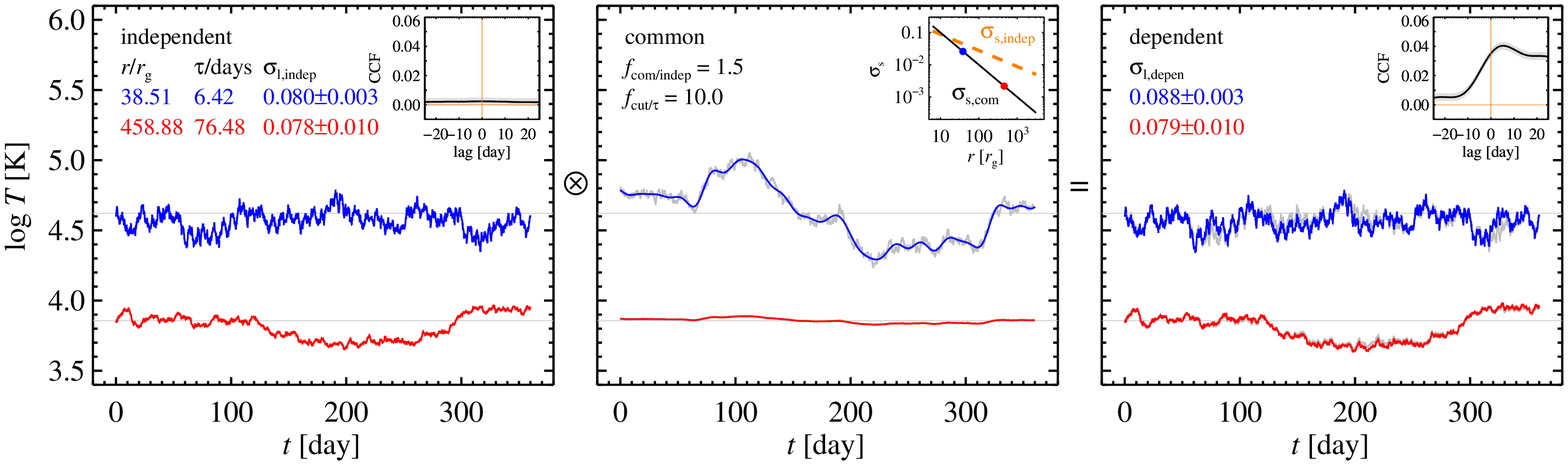}
\caption{
An illustration of constructing the {\it dependent} temperature fluctuations (right panel) by coupling the smoothed {\it common} large-scale temperature fluctuation (middle panel) with the {\it independent} ones (left panel or the oscillating gray lines in the right panel) in two arbitrary regions at $r = 38.51~r_{\rm g}$ (blue lines) and $458.88~r_{\rm g}$ (red lines) with local damping timescales $\tau = 6.42$ days and 76.48 days, respectively.
In all panels, the two horizontal gray lines give the corresponding temperatures of the standard thin disk at these two radii.
The independent temperature fluctuations (the oscillating blue and red lines in left panel) are described by the CAR(1) process with variation on short timescale of $\sigma_{\rm s,indep}(r) = \sigma_{\rm s,indep}(r_{\rm in}) \times (r/r_{\rm in})^{-0.5}$ (cf. the orange dashed line in the inset of middle panel), while the common fluctuations (the oscillating gray lines in middle panel) are described by a pure random walk process with variation on short timescale of $\sigma_{\rm s,com}(r) \equiv f_{\rm com/indep} \times \sigma_{\rm s,indep}(r_{\rm in}) \times (r/r_{\rm in})^{-1}$ (cf. the black solid line in the inset of middle panel), and further smoothed by cutting off high frequencies below $\tau_{\rm cut} = f_{\rm cut/\tau} \times \tau(r_{\rm in})$ (i.e., the smoothed blue and red lines in middle panel).
After including the common fluctuations, the dependent temperature variation on long timescale is slightly increased, i.e., $\sigma_{\rm l,depen} \gtrsim \sigma_{\rm l,indep}$, averaging over $N_\phi$ regions at a given radius. More important and interesting is the emergence of a small correlation and a positive lag of several days, after averaging correlations of $N^2_\phi$ pair of the dependent temperature fluctuations between smaller and larger radii (the black line with the $3\sigma$ error as gray region in the inset of right panel compared to that in the inset of left panel). The {\it dependent} temperature fluctuations at larger radii statistically lag those at smaller radii is as a result of {\it the quicker (i.e., shorter local damping timescale) regression capability at smaller radii than at larger ones} when responding to the common large-scale fluctuation.
Refer to Section~\ref{sect:model} for more details.
}\label{fig:tempfluc_indep2depen}
\end{figure*}

With the aforementioned difficulties for the reprocessing scenario in mind, we are thinking about whether the lag is not originated from the reprocessing. Further considering the success of the inhomogeneous accretion disk scenario on explaining several observational aspects, we start to explore the possibility that whether the lag can directly stem from the disk within the inhomogeneous scenario and without light echoing.

We consider the inhomogeneous accretion disk scenario originally proposed by \citet{DexterAgol2011}, later justified by \citet{Ruan2014}, and recently enriched by \citet{Cai2016}. 
In analog to \citet{Cai2016}, the accretion thin disk is divided into 4356 square-like zones in $r$ and $\phi$ with $N_{\rm r} = 66$ layers and $N_\phi = 66$ zones per each layer, from $r_{\rm in} = 6~r_{\rm g}$ to $r_{\rm out} = r_{\rm in} f^{N_{\rm r}}_{\rm rbr} \simeq 3200~r_{\rm g}$, where the gravitational radius $r_{\rm g} \equiv GM_\bullet/c^2$ and the radial boundary ratio of each layer $f_{\rm rbr} = 1.1$. Fewer zones were considered in \citet[][i.e., $N_\phi = 24$ and $f_{\rm rbr} = 1.3$]{Cai2016} in order to mainly save the computational times while keeping the conclusions. The new selection of zone number is to further enhance (although minor) the correlation of light curves among UV and near-infrared bands once a common fluctuation has been considered over the whole disk. The new disk split has $\sim 480$ zones per factor of two in radius, which is also more consistent with that suggested by \citet{DexterAgol2011}. The new outer radius of disk is still large enough for the relevant longest wavelength of NGC 5548 considered here.

The classical temperature profile of standard thin disk, i.e., $T_{\rm d} \propto r^{-3/4}$, is assumed and viewed face-on as well as zero spin of the black hole. Besides these, the relativistic effects and disk atmosphere radiative transfer are neglected at this stage. 
Thereafter, the fist-order continuous autoregressive (CAR(1)) process\footnote{The CAR(1) process is firstly introduced by \citet{Kelly2009} from statistics to the AGN variability community. It is often referred to as a damped random walk process in the astronomical literature \citep[e.g.,][]{Kozlowski2010,Macleod2010,Andrae2013,Zu2013}, while to an Ornstein-Uhlenbeck process in the physics literature. This process is just a simple case of the general class of continuous-time autoregressive moving average models \citep{Kelly2014}.} \citep{BrockwellDavis2002book,Kelly2009} is assumed to depict the independent temperature fluctuation in logarithm for each zone, where the logarithmic temperature, $\log T_{\rm indep}(t, r, \phi)$, with a long-term variance of $\sigma^2_{\rm l,indep}$, returns to the mean, $\log T_{\rm return}(r) = \log T_{\rm d}(r) - 2 \sigma^2_{\rm l,indep} \ln(10)$, at a radius-dependent characteristic damping timescale $\tau(r)$ \citep[see Section 2 of ][for more details]{Cai2016}. 

As illustrated in Figure~\ref{fig:tempfluc_indep2depen}, further improvement in this contribution is the consideration of a speculated {\it common} large-scale temperature fluctuation, $\log T_{\rm com}(t, r)$ (cf. its middle panel), being coupled with different local {\it independent} temperature fluctuations, $\log T_{\rm indep}(t, r, \phi)$ (cf. its left panel), and ending up with {\it dependent} temperature fluctuations, $\log T_{\rm depen}(t, r, \phi)$ (cf. its right panel).

More mathematically, for a region around $r$ and $\phi$, its {\it independent} temperature fluctuation, probably initialized by disk instabilities (e.g., the inflow/viscous/secular instability by \citealt{LightmanEardley1974}; the thermal instability by \citealt{ShakuraSunyaev1976}; the magneto-rotational instability by \citealt{BalbusHawley1991}; the photon bubble instability by \citealt{Turner2005}), is empirically modeled as a CAR(1) process by \citep[cf.][]{DexterAgol2011}
\begin{align}\label{eq:indep}
	d \Delta \log T_{\rm indep}(t, r, \phi) = &- \frac{\Delta \log T_{\rm indep}(t, r, \phi)}{\tau(r)} dt \nonumber \\
	&+ \sigma_{\rm s,indep}(r) \sqrt{dt} \epsilon(t, r, \phi),
\end{align}
where $\Delta \log T_{\rm indep}(t, r, \phi) \equiv \log T_{\rm indep}(t, r, \phi) - \log T_{\rm return}(r)$ is the temperature deviation from the mean temperature $\log T_{\rm return}(r)$ at epoch $t$, the variation per day is $\sigma^2_{\rm s,indep}(r) = 2 \sigma^2_{\rm l,indep}/\tau(r)$, and $\epsilon(t, r, \phi)$ is Gaussian white noise process with zero mean and variance equal to 1. 
As a reference, we assume a constant $\sigma_{\rm l,indep} = 0.08$ dex and radius-dependent $\tau(r) = \tau_0 (r/r_{\rm in})^\alpha$, where $\tau_0 \equiv \tau(r_{\rm in}) = 1.0$ day and $\alpha = 1.0$, for NGC 5548. 
Therefore, $\sigma_{\rm s,indep}(r) = \sigma_{\rm s,indep}(r_{\rm in}) \times (r/r_{\rm in})^{-\alpha/2}$ and $\sigma_{\rm s,indep}(r_{\rm in}) = \sigma_{\rm l,indep} \sqrt{2/\tau_0} = 0.113$.
Comparing to the adopted parameters in \citet[][i.e., $\sigma_{\rm l,indep} = 0.2$ dex, $\tau_0 = 10$ days, and $\alpha = 1.0$ for $M_\bullet = 5 \times 10^8 M_\sun$]{Cai2016}, the smaller $\sigma_{\rm l,indep}$ is mainly constrained by the UV/optical variation amplitudes of NGC 5548 and it is not so important on determining the correlation strength and lag (see discussions below). The $\tau_0$ results from a simple scaling for the black hole mass of NGC 5548 and $\alpha$ is fixed to be same as before since no strong constraints can be inferred at the moment.

The propagation and mixing of fluctuations induced by disk instabilities at different timescales may lead to a net smoothed common large-scale fluctuation across a wide disk region beyond a characteristic timescale, which may be related to an average speed of propagations. The fluctuations over a large disk region are common in the sense that their shapes but not amplitudes of the power spectrum density beyond a characteristic timescale are similar, and they do not necessarily happen simultaneously over different regions. Due to the complexity of the poorly understood propagations, we firstly assume
the {\it common} large-scale temperature fluctuation to be a pure random walk process (or a damping timescale of $\tau \rightarrow \infty$ for the CAR(1) process) and described by
\begin{equation}\label{eq:comm}
	d \log T_{\rm com}(t, r) = \sigma_{\rm s,com}(r) \sqrt{dt} \epsilon'(t),
\end{equation}
where $\epsilon'(t)$ is another Gaussian white noise process with zero mean and variance equal to 1, and the variation per day is $\sigma^2_{\rm s,com}(r)$.
Being constrained by the observed anti-correlation between the variation amplitude and the UV/optical wavelength, the temperature fluctuations on short timescale are generally required to be weaker with increasing radius. By analogy with the relation for $\sigma_{\rm s,indep}(r) \propto r^{-\alpha/2}$, we may model the variation amplitude of common fluctuation on short timescale 
as $\sigma_{\rm s,com}(r) = f_{\rm com/indep} \times \sigma_{\rm s,indep}(r_{\rm in}) \times (r/r_{\rm in})^\gamma$, where $f_{\rm com/indep} = 1.5$ is the assumed relative contribution between common and independent fluctuations on short timescale at $r_{\rm in}$, and the index $\gamma$ describes the decreasing trend. The $\gamma = - \alpha / 2$ indicates the same decreasing trend for both $\sigma_{\rm s,com}$ and $\sigma_{\rm s,indep}$ (e.g., $\gamma = -0.5$ if $\alpha = 1.0$). For $\alpha = 1.0$, our current selection of $\gamma = -1.0$ results in a faster decreasing trend for $\sigma_{\rm s,com}$ to avoid too large variation amplitudes at long wavelengths dominated by radiation of outer radii.
Secondly, the high-frequency portions of the common fluctuation below a characteristic timescale of $\tau_{\rm cut}$ are smeared out in order to capture the smooth property of the mixing effect of many distinct outward/inward propagations over the disk. This characteristic timescale is expected to be inversely proportional to the average speed of propagations because for a disk region slower propagation of fluctuation means the whole region could experience the same fluctuation only beyond a longer smoothing timescale.
Here, we assume a constant $\tau_{\rm cut} = f_{\rm cut/\tau} \times \tau(r_{\rm in})$ and $f_{\rm cut/\tau} = 10.0$.
Note that although the variance of a random walk (not damped) is infinite, the mean value of $\log T_{\rm com}$ is less relevant because the one imprinted on the independent fluctuations is only the difference, $\delta \log T_{\rm com}(s, r) \equiv \log T_{\rm com}(t, r) - \log T_{\rm com}(s, r)$, between two adjacent epochs, e.g., $t > s$ and $\delta t = t - s$. 

To construct the dependent temperature fluctuation, we tentatively modify the expressions for CAR(1) process. We imagine that, from epochs $s$ to $t$, the departure of local temperature fluctuation at epoch $s$, $\Delta \log T_{\rm depen}(s, r, \phi)$, is firstly shifted by the common large-scale fluctuation through its difference, $\delta \log T_{\rm com}(s, r)$, before they would together return back somewhat to the mean level at a local damping timescale $\tau(r)$, that is,
\begin{align}\label{eq:depen_integral}
	&\Delta \log T_{\rm depen}(t, r, \phi) = [\Delta \log T_{\rm depen}(s, r, \phi) + \delta \log T_{\rm com}(s, r)] \nonumber\\ 
	&\times \mbox{e}^{-\delta t/\tau(r)} + \sqrt{\frac{\tau(r) \cdot \sigma^2_{\rm s,indep}(r)}{2} [1 - \mbox{e}^{-2 \delta t/\tau(r)}] } \times \epsilon(s, r, \phi),
\end{align}
where $\Delta \log T_{\rm depen}(t, r, \phi) \equiv \log T_{\rm depen}(t, r, \phi) - \log T_{\rm return}(r)$ is the dependent temperature deviation from the mean of independent fluctuations.
Note that here ``$\Delta$'' denotes departure from the mean level at a single epoch, while ``$\delta$'' the difference between two adjacent epochs.
Equivalently, the {\it dependent} temperature fluctuation in differential form is represented by
\begin{align}\label{eq:depen}
	d \Delta \log &T_{\rm depen}(t, r, \phi) = - \frac{\Delta \log T_{\rm depen}(t, r, \phi) }{\tau(r)} dt \nonumber \\
	&+ d \log T_{\rm com}(t, r) + \sigma_{\rm s,indep}(r) \sqrt{dt} \epsilon(t, r, \phi).
\end{align}
If there is no common large-scale fluctuation, the above equation would be identical to that for independent fluctuation, i.e., Equation~(\ref{eq:indep}).
In the following comparisons, we have used the same white noise process, $\epsilon(t, r, \phi)$, at given radius and azimuth for both dependent and independent fluctuations in order to easily highlight the effect caused by considering the common large-scale fluctuation without being annoyed by additional random process.

\begin{figure*}[tb!]
\centering
\includegraphics[width=0.98\textwidth]{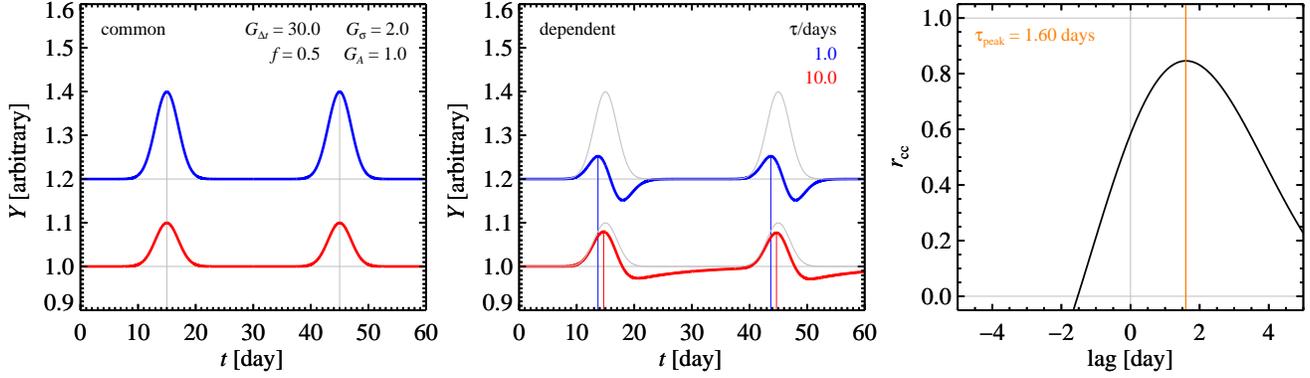}
\caption{An illustrated scheme for the emergence of lag. The independent temperature fluctuation is assumed to be negligible (i.e., $\sigma_{\rm s,indep} \rightarrow 0$ but a finite damping timescale $\tau$), while the common fluctuation is represented by a series of Gaussian pulses, whose standard deviation is $G_{\sigma} = 2.0$, area is $G_{\rm A} = 1.0$, and separated by $G_{\Delta t} = 30.0$ days. In the left panel, the blue and red lines show the initial common fluctuation and its rescaled version, by a factor of $f = 0.5$, respectively, while the vertical light-gray lines indicate the peak epochs of those pulses. 
The middle panel illustrates the dependent fluctuations when the corresponding common fluctuations are returned to their ground states at different damping timescales, i.e., 1.0 and 10.0 days for the blue and red ones, respectively. For comparison, the light-gray pulses in the middle panel indicate the corresponding common fluctuations (or the dependent fluctuations when damping timescales are infinite), while the thin vertical blue and red lines nominate the peak epochs of the blue and red lines, respectively. The difference of these vertical lines intuitively illustrates the lag between the blue and red lines, which is also quantitatively presented in the right panel following the analysis of cross-correlation.
}\label{fig:test_indep2depen}
\end{figure*}

After enough ``burn-in'' time for each zone, i.e., longer than its characteristic damping timescale, we have simulated the temperature fluctuations in steps of $\Delta t = 0.1$ days up to a time length of 9.7 years. The maximal time length is chosen such that we would have 20 segments of temperature fluctuation and each segment has a time length in the observed frame comparable to that of the {\it Hubble Space Telescope} ({\it HST}) light curves of NGC 5548, i.e., $\simeq 180$ days.
Figure~\ref{fig:tempfluc_indep2depen} only illustrates the temperature fluctuations with a time length of about two segments in two arbitrary regions around $r = 38.51~r_{\rm g}$ (blue lines) and 458.88 $r_{\rm g}$ (red lines). 
For our assumed black hole mass and Eddington ratio, these two small and large radii are selected owing to their dominant contributions to the wavelengths of $\sim 1000~\angstrom$ and $\sim 9000~\angstrom$, respectively, which are almost the minimal and maximal wavelengths concerned here.
As expected and presented as the black line in the inset of left panel of Figure~\ref{fig:tempfluc_indep2depen}, the independent temperature fluctuations are statistically uncorrelated and of zero lag when averaging correlations of $N^2_\phi$ pair of temperature fluctuations at the two selected radii. This is also true for any other two radii and it is simply the implication of the inhomogeneous disk models with totally independent temperature fluctuations. 
Note that the 3 $\sigma$ uncertainty of the mean correlation (gray regions in the insets of left and right panels of Figure~\ref{fig:tempfluc_indep2depen}) is crudely estimated as the 3 $\sigma$ standard deviation of $N^2_\phi$ pair of correlations divided by $\sqrt{N^2_\phi}$. 

Having included the common large-scale fluctuations, the dependent temperature variation on long timescale is slightly increased, i.e., $\sigma_{\rm l,depen} \gtrsim \sigma_{\rm l,indep}$, averaging over $N_\phi$ regions at a given radius.
In both two cases, the simulated temperature fluctuation within a zone at a single epoch may have an effective Eddington ratio, $\lambda_{\rm Edd,eff}(t) \equiv \lambda_{\rm Edd} T^4(t)/T^4_{\rm d}$, greater than unity even with a quite small probability of $<0.14\%$. These instant epochs may represent the process through which the energy is quickly released after a relatively long accumulating period. However, more relevant is the Eddington ratio corresponding to the bolometric luminosity integrating over all zones at a single epoch. 
Being assessed with ten times of simulations, for the independent case, the single-epoch Eddington ratios inferred from the bolometric luminosity are very close to the adopted value and their mean is 0.099 with a tiny scatter of 0.002. Instead, for the dependent case, the mean Eddington ratio is 0.107 with a scatter of 0.040, and the maximal instant Eddington ratio can only be as large as $\sim 0.4$.

It is also not so intuitive to notice that the correlation has been enhanced, especially when directly comparing the dependent temperature fluctuations (the blue and red lines in the right panel of Figure~\ref{fig:tempfluc_indep2depen}) with the independent ones (the gray oscillating lines in the same panel).
Instead, after averaging over correlations of $N^2_\phi$ pair of the dependent temperature fluctuations at the two radii, a non-zero correlation yet small and a positive lag of several days emerge statistically. 
Note that the correlation also behaves asymmetric toward its lagged direction.

Although the strength of correlation between dependent temperature fluctuations at the two selected radii is small, it would be further increased when increasing the relative contribution of the common fluctuation, i.e., increasing $f_{\rm com/indep}$ and/or larger $\gamma$. However, further enhancement on the correlation may lead to too large UV/optical variation amplitudes. Actually, the adopted values for $f_{\rm com/indep}$ and $\gamma$ are limited by the UV/optical variation amplitudes of NGC 5548 (see next section). Another parameter that would affect the UV/optical variation amplitudes is $\sigma_{\rm l,indep}$. However, it does little effect on the correlation and lag since by changing it only a simple scaling of the amplitudes of temperature fluctuation would be performed.

The emergence of a positive lag and the asymmetry is the result of {\it a quicker regression capability at smaller radii (i.e., shorter local damping timescale) than at larger ones} when responding to the same common large-scale fluctuation. 
For better understanding the lag, an illustrated scheme is presented in Figure~\ref{fig:test_indep2depen} where the independent fluctuation has been neglected (i.e., $\sigma_{\rm s,indep} \rightarrow 0$ but a finite damping timescale $\tau$) and the common fluctuation is simplified as a series of Gaussian pulses. Then, following Equation~(\ref{eq:depen}), the effect of the common fluctuation is added to an arbitrary constant and is returned to the constant at different damping timescales. With a shorter damping timescale (e.g., the blue one with 1.0 day), the fluctuation reaches its peak more quickly or more earlier than that affected by a longer damping timescale (e.g., the red one with 10.0 days). Since the smaller damping timescale the larger regression capability of fluctuation, we therefore attribute the lag to the {\it differential regression capability}.
Consequently, the lag would increase with increasing the ratio difference of damping timescales between the larger and smaller radii or with increasing $\alpha$. Moreover, the lag also increases with slightly increasing the smoothing timescale $\tau_{\rm cut}$ in a quite sensitive way. This is because the smoothed common large-scale fluctuations become more monotonic increasing/decreasing or the period of increasing/decreasing becomes longer when $\tau_{\rm cut}$ is larger. However, the dependence would become more complicated because too large $\tau_{\rm cut}$ would give rise to very small variation amplitudes of the smoothed common large-scale fluctuations and, therefore, the independent fluctuations would dominate over the smoothed common ones, leaving poor correlation and unidentifiable lag. A third trivial way that would increase lag is increasing $\tau_0$ because this is also a simple scaling of timescale.

In sum, there are about six parameters in the model that would be related to both/either the UV/optical variation amplitudes and/or the lag-wavelength relation. Currently, we focus on presenting the main idea for intrinsically generating a lag-wavelength relation of UV/optical continua in this upgraded inhomogeneous accretion model with a reasonable selection of these parameters, while an exhaustive investigation of how these parameters would determine the UV/optical variation amplitudes and the lag-wavelength relation as well as several other properties, e.g., the structure function and color variations, of the simulated UV/optical continua would be postponed to an upcoming paper in this series (Z. Y. Cai et al. 2018, in preparation).

\section{Tight inter-band correlations and small lags}\label{sect:results}

\begin{figure*}[tb!]
\centering
\includegraphics[width=0.98\textwidth]{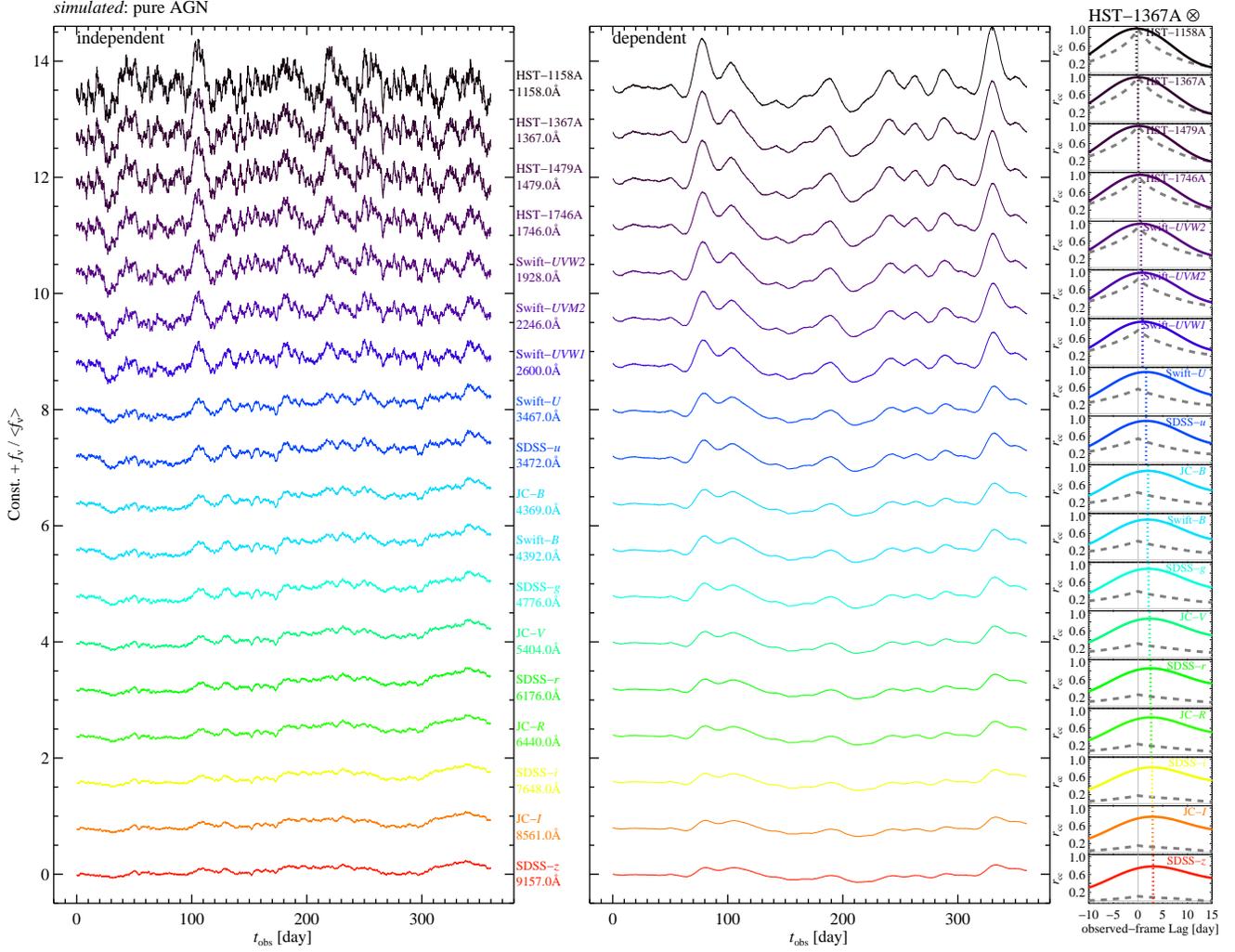}
\caption{An illustration of the simulated continuum light curves in observed frame from UV (top) to $z$ (bottom) bands for the accretion thin disk (pure AGN) with the {\it independent} (left panel) and {\it dependent} (middle panel) temperature fluctuations.
The light curves are shifted horizontally for clarity and those from the independent model are further rescaled to the same variation amplitudes as the corresponding ones from the dependent model.
The right panels illustrate the cross-correlations between the {\it HST} 1367 $\angstrom$ light curve and the other ones (nominated at the top-right corner of each panel) for the independent (gray dashed lines) and dependent models (colored solid lines).
Prominently, the {\it dependent} model gives impressive results that {\it the UV/optical light curves are highly coordinated} and that {\it inter-band lags exist between the light curves}, relative to the 1367 $\angstrom$ one, as indicated by the vertical dotted lines for the lag at correlation peak. 
}\label{fig:lc_indep_vs_depen}
\end{figure*}

In the last section, we have introduced both the independent and dependent temperature fluctuations, and elucidated the emergence of a weak correlation and a lag between the dependent temperature fluctuations at different radii. 
Since the light curves are obtained by integrating the temperatures over the entire disk after complicated manipulations, it may be not so intuitive to get insight on the inter-band lags between light curves. 
However, if there is no lag between temperature fluctuations at different radii, it would be difficult to expect any inter-band lag between light curves. Therefore,
to examine whether there is also a lag between light curves and further to better compare with the observed light curves and their statistical properties, we need to go from temperature fluctuations to the corresponding fluctuations at given wavelengths.

At any epoch $t$, summing up all the assumed blackbody emissions of each zone, we would obtain the corresponding fluctuating spectral energy distributions, which are redshifted to the observed frame and then convolved with the transmission curve at a given band for the corresponding light curve. 
Following \citet[][cf. their Table~5]{Fausnaugh2016} for NGC 5548, we consider the light curves in 18 bands, including {\it HST} (HST-1158 $\angstrom$, -1367 $\angstrom$, -1479 $\angstrom$, -1746 $\angstrom$), {\it Swift} (Swift-$UVW2,UVM2,UVW1,U,B$), Johnson/Cousins (JC-$BVRI$), and Sloan Digital Sky Survey (SDSS-$ugriz$). 

Firstly, the ideal evenly-sampled light curves implied by accretion disk (or pure AGN) and relevant standard cross-correlation functions (CCFs) are compared for both the independent and dependent models in Section~\ref{sect:ideal_model}. Followed in Section~\ref{sect:ngc5548} is the more delicate comparisons of the unevenly-sampled light curves and relevant linearly interpolated cross-correlation functions (ICCFs) for both the Seyfert galaxy NGC 5548 and the dependent model, including host contamination as well as taking into account the sampling and photometric error.

\subsection{An ideal case: two distinguishable models}\label{sect:ideal_model}

The evenly-sampled light curves, $f_\nu(t, \lambda)$, in the observed frame from UV to $z$ bands are shown in Figure~\ref{fig:lc_indep_vs_depen} for the accretion thin disk (or pure AGN) with the independent (left panel) and dependent (middle panel) temperature fluctuations. The shown light curves are divided by their own mean, $\langle f_\nu(t, \lambda) \rangle_t$, and are shifted horizontally for clarity.
Also shown from up to down in the right panels of the same figure are the CCFs between the evenly-sampled 1367 $\angstrom$ light curve and the other ones, nominated at the top-right corner of the corresponding panel, for the independent (gray dashed lines) and dependent models (colored solid lines).
Note that only the same temporal portion of the simulated light curves as in Figure~\ref{fig:tempfluc_indep2depen} for temperature fluctuations is shown, while the cross-correlations are calculated linearly, i.e., with $f_\nu$, using the whole simulated light curves. We confirm that the resulting CCFs are almost the same if using logarithmic flux, $\log f_\nu$.

As discussed in the last section and shown in Figure~\ref{fig:tempfluc_indep2depen}, the long-term variation of dependent temperature fluctuation is slightly larger than that of independent one for each individual zone. As a result, the variations of light curves implied by the dependent model are also larger, by factors of $\sim 7-18$ increasing with decreasing wavelength, than those implied by the independent model. Therefore, the light curves implied by the latter are rescaled (actually enlarged) to the same variation amplitudes as those of the former for a clear comparison in Figure~\ref{fig:lc_indep_vs_depen}. 
Note that the variation of temperature fluctuation does not largely increase from the independent model to the dependent model for each individual zone, but the implied variations of light curves do change a lot. This is because after summing up the radiation variations converted from the temperature fluctuations of multiple zones, the variations of light curves from the independent model are depressed due to the cancellation of completely independent fluctuations, while the variations of light curves from the dependent model are instead enhanced due to the common fluctuation over all zones. The difference of variation amplitude between the independent model and the dependent model is related to the total number of zones in such a way that more zones larger difference.

Due to the radiation overlapping at given wavelength from different but close regions, the independent model (cf. the left panel and the gray dashed lines in the right panels of Figure~\ref{fig:lc_indep_vs_depen}) also implies quite strong correlations among the light curves from the HST-1158$\angstrom$ to Swift-$UVW1$ or from the SDSS-$u$ to SDSS-$z$. However, the correlation becomes very weak when the wavelength difference is large enough, e.g., between $\sim 1000~\angstrom$ and $\sim 9000~\angstrom$ for the extreme case. Although the correlations of light curves at close wavelengths could be large, the corresponding CCFs possess a cusp near their peaks, owing to the quick oscillation of light curve. Moreover, there is completely non-lag among the UV/optical light curves. All these implications are at odds with observations.

Instead, the UV/optical light curves implied by the dependent model are impressively coordinated, even for the extreme wavelength difference (cf. the middle panel and the colored solid lines in the right panels of Figure~\ref{fig:lc_indep_vs_depen}). More important and interesting is the emergence of noticeable longer lags with increasing the wavelengths relative to the 1367 $\angstrom$ (cf. the vertical dotted lines in the right panels of Figure~\ref{fig:lc_indep_vs_depen} for the lag at correlation peak). 

Firstly, concerning with the light curves of the dependent model, the highly correlated peaks and troughs over all the wavelengths are directly imprinted by the assumed {\it common} large-scale fluctuation, which exactly shows the same peaks and troughs near similar epochs (cf. the middle panel of Figure~\ref{fig:tempfluc_indep2depen}). 
In other words, the light curves implied by the dependent model at different wavelengths behave as a common low-frequency variation plus different high-frequency ones with stronger power at shorter wavelength.

Besides these, the characteristic timescales of light curves at all wavelengths implied by the dependent model seems to be similar with each other and longer than those of the corresponding light curves implied by the independent model, where the characteristic timescales have been shown to decrease with decreasing wavelength by fitting the structure function of the CAR(1) process to that of the simulated light curve (cf. the bottom panel of Figure~6 in \citealt{Cai2016}). 
In \citet{Cai2016}, the characteristic timescales of light curves are determined by the local damping timescales of temperature fluctuations at disk regions dominating the radiation at relevant wavelengths. Since the local damping timescales of temperature fluctuations increase with increasing radius, i.e., $\tau(r) = \tau_0 (r/r_{\rm in})^\alpha$, and radiation at longer wavelength is primarily from outer disk regions, the implied timescales of light curves would also increase with increasing wavelength. 
However, the UV/optical light curve implied by the dependent model is now a manifestation of the common large-scale temperature fluctuation, which is characterized by the smoothing timescale $\tau_{\rm cut}$. This parameter is currently assumed to be constant over all radii and this may be the reason for a similar characteristic timescale of UV/optical light curves implied by the dependent model. On further comparing with observations and exploring any possible radius-dependence of $\tau_{\rm cut}$, they will be discussed in a separate paper of this series (Z. Y. Cai et al. 2018, in preparation).

On the shape of CCFs implied by the dependent model (cf. the colored solid lines in the right panels of Figure~\ref{fig:lc_indep_vs_depen}), they are quite smooth around their peaks, consequently, more consistent with observations. Another noticeable property of the CCFs of light curves at longer wavelengths, e.g., $\sim 9000~\angstrom$, relative to the 1367 $\angstrom$, is that their primary peak trends to be a little asymmetric and skewed toward larger lags. 
This asymmetry of correlation between light curves in different bands is similar to that between temperature fluctuations at different radii as we have introduced in Section~\ref{sect:model} and Figure~\ref{fig:tempfluc_indep2depen}.
This property is of course a testable prediction of the dependent model.

\subsection{A real case: NGC 5548}\label{sect:ngc5548}

\begin{figure*}[tb!]
\centering
\includegraphics[width=0.98\textwidth]{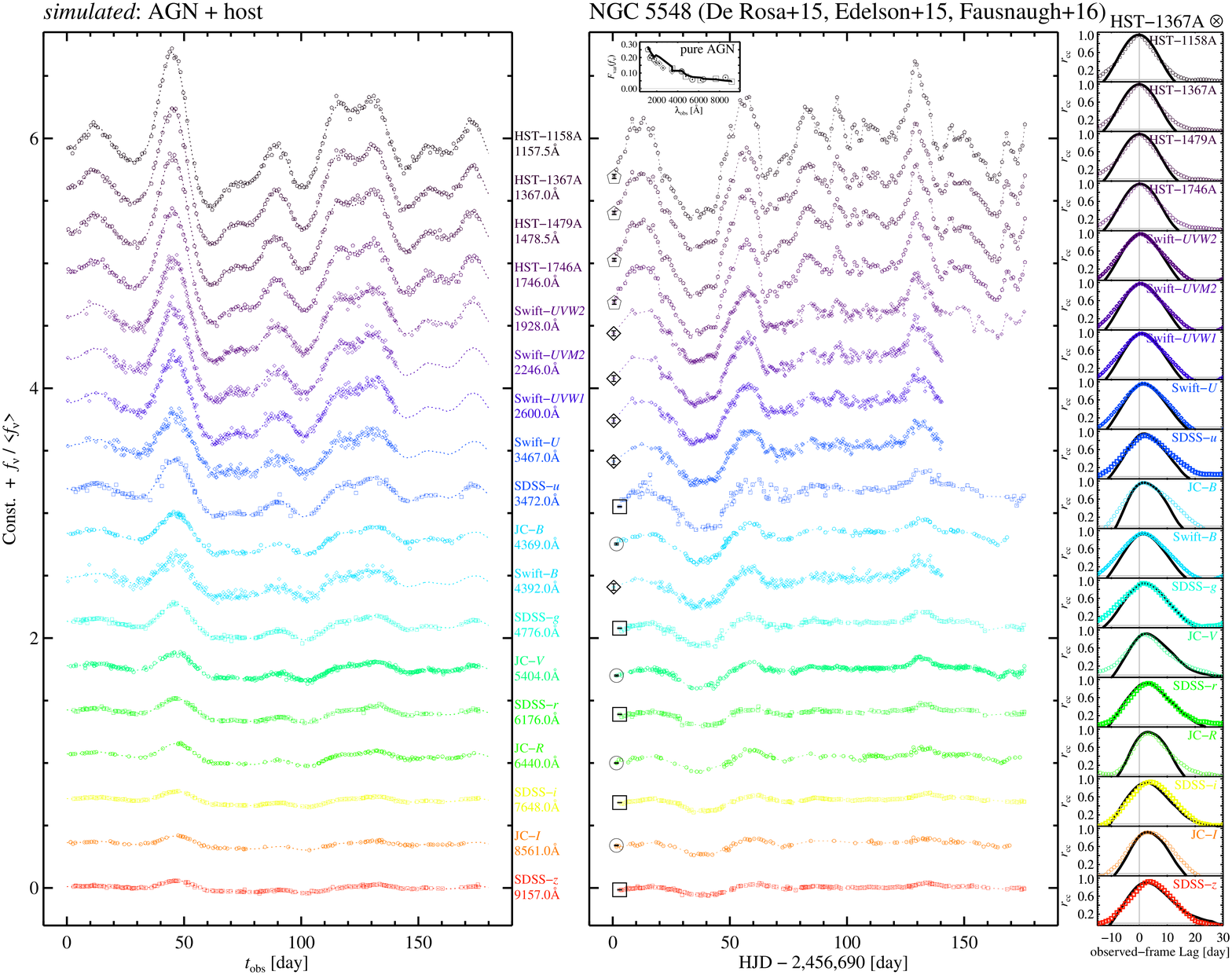}
\caption{An illustration of a selected segment of light curves (left panel) in observed frame from UV (top) to $z$ (bottom) bands, whose time length as well as the other statistical properties are all in analogy with those of NGC 5548 (middle panel).
In the left panel, the initial simulated light curves are plotted as dotted lines, while those after sampling and assigning errors are scattered as symbols around the dotted ones with increasing wavelength from up to down ({\it HST} as pentagons, {\it Swift} as diamonds, Johnson/Cousins as circles, and SDSS as squares).
In the middle panel, the UV/optical data of NGC 5548 are taken from \citet[][{\it HST}]{DeRosa2015}, \citet[][{\it Swift}]{Edelson2015}, and \citet[][Johnson/Cousins and SDSS]{Fausnaugh2016} with median errors indicated at the beginning of the light curves.
The inset at the top of the middle panel shows the fractional variabilities, $F_{\rm var}$, as a function of wavelength for the selected segment of UV/optical light curves (black line for pure AGN emission) and the estimations by \citet[][symbols for those corrected for host contamination]{Fausnaugh2016}.
In the right panels, the linearly interpolated cross-correlation functions of these UV/optical light curves with increasing wavelength relative to the 1367 $\angstrom$ one are shown from up to down for the selected segment (black lines) and those estimated for NGC 5548 (colored symbols).
}\label{fig:lc_depen_vs_ngc5548}
\end{figure*}

To further quantitatively compare the simulated light curves in flux and their CCFs with those of NGC 5548, we equally divide the whole simulated light curve at given wavelength into 20 segments, each of which is in analogy of NGC 5548 with an observed time length of $\sim 180$ days, considered by \citet{Fausnaugh2016}. 
We also independently run the simulation ten times and therefore we have totally 200 segments of light curves.
Furthermore, a constant host emission at given wavelength has been added to the segmented light curve of pure AGN. 
Note that the consideration of host contribution solely aims to directly compare the variation amplitudes of the simulated and observed light curves, while the results of correlation functions as well as the color variations do not rely on the assumed host emission.
The fractions of host contamination are assumed using values estimated by \citet[][see their Table~5]{Fausnaugh2016}, only available for wavelengths longer than $\sim 3400~\angstrom$ or {\it Swift} $U$ band.
Thereafter, the segmented light curves are re-sampled by linear interpolation at the same epochs as NGC 5548, and then assigned with flux uncertainties randomly selected from Gaussian deviations with the mean and standard deviation of measurement errors in percentage of NGC 5548 at corresponding wavelength. Finally, the fluxes are fluctuated according to Gaussian deviations with zero mean and standard deviation equaling to the former assigned uncertainty.
Note that no further operation has been applied to these segments until we consider their color variations in Section~\ref{sect:color_variation}.

As shown in Figure~\ref{fig:lc_depen_vs_ngc5548} for one of the 200 segments, selected by its similarity to NGC 5548 in several properties discussed hereafter, the variation amplitudes of segmented light curves including host contaminations are intuitively comparable to those of real light curves of NGC 5548 (cf. the left and middle panels). In the meantime, the fractional variabilities \citep[see Equations 2-3 in][]{Fausnaugh2016} of the re-sampled light curves contributed by pure AGN as a function of wavelength (cf. the black line in the inset of middle panel) are quantitatively in agreement with those estimations by \citet{Fausnaugh2016} after having corrected for host-galaxy flux (cf. the symbols in the inset of middle panel). 

The right panels of Figure~\ref{fig:lc_depen_vs_ngc5548} from up to down illustrate the ICCFs, as employed by \citet{Peterson1998,Peterson2004}, of these UV/optical light curves with increasing wavelength relative to the 1367 $\angstrom$ one, for both the re-sampled light curves (the black solid lines) and the real data of NGC 5548 (the symbol-like lines). When calculating the ICCF, the 1367 $\angstrom$ light curve is shifted in time on a grid of lags $\tau$ by 0.01 days and then linearly interpolated to the same epochs of the other light curve before calculating the correlation coefficient $r_{\rm cc}(\tau)$.
Except that re-sampling and assigning photometric errors have been employed onto the simulated initial light curves, no further detrending procedure has been applied in order to reduce as much as possible the uncertainties introduced by any extra operation on light curves. Then we also identically calculate the ICCFs for the real data of NGC 5548 and we confirm that both the ICCFs and their properties, e.g., maximal correlation, peak and centroid lags (see the definitions in the next section), are all qualitatively consistent with those estimated by \citet{Fausnaugh2016}. 

Generally, the agreement between the ICCFs of the selected segment of simulation and those of NGC 5548 is encouraging. Of course, there are differences between those un-selected segments of simulation and NGC 5548. However, these differences are easily expected and it is really difficult to establish a complete match between the simulation with random processes and the observation on a single source. Nevertheless, the consistence between simulation and NGC 5548 is still justified by the fact that a nice statistical agreement of the ICCF properties of simulation and NGC 5548 can be established as discussed in the following.

\section{Discussions}\label{sect:discussion}

In the last section, we have introduced a specific comparison on the properties of UV/optical light curves between a preferred segment selected out of 200 simulated segments and NGC 5548. Now, we start to statistically compare the ICCF/CCF properties (see Sections~\ref{sect:coordination}-\ref{sect:width-wave}), including their maximal correlation ($r_{\rm cc,max}$ for the peak of ICCF/CCF), the peak lag ($\tau_{\rm peak}$ for the lag at the peak of ICCF/CCF), the centroid lag ($\tau_{\rm cent}$ for the correlation-weighted mean lag with $r_{\rm cc}(\tau) > 0.8~r_{\rm cc,max}$), and the width ($W_{\rm cent} \equiv \tau_{\rm max} - \tau_{\rm min}$ at $r_{\rm cc}(\tau) = 0.8~r_{\rm cc,max}$), as well as the color variation (see Section~\ref{sect:color_variation}), between all 200 segments and NGC 5548 over the UV/optical wavelengths.

Although \citet{Fausnaugh2016} have already reported the maximal correlation, the peak lag, and the centroid lag of ICCFs for NGC 5548, we have re-calculated these ICCF properties for two reasons.
The primary one is that we would like to compare the width of ICCF, which is unavailable in \citet{Fausnaugh2016}, since it contains the information on the fluctuation quickness, i.e., quicker fluctuation smaller width.
The second one is that we have not applied a detrending procedure to the light curves in order to avoid introducing further uncertainties. For example, \citet{Fausnaugh2016} have applied a detrending procedure by subtracting a second-order polynomial linear least-squares fit from the observed light curves, and the resultant lags of the {\it Swift} light curves are systematically larger than those found in \citet{Edelson2015} using a 30-day running mean.
Therefore, for the 1367 $\angstrom$ light curve relative to the other one, we derive these properties for both observation and simulation.

\begin{figure}[tb!]
\centering
\includegraphics[width=\columnwidth]{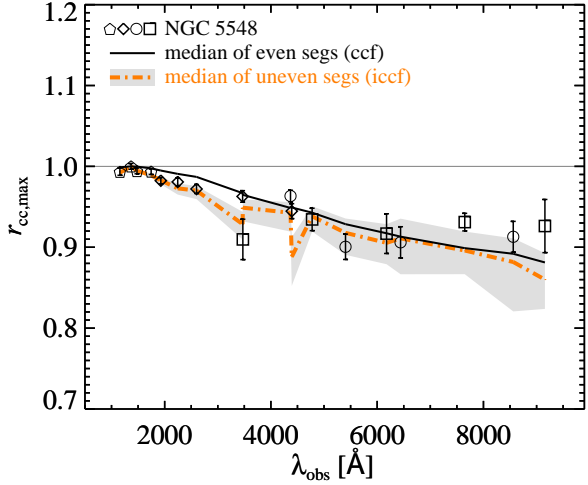}
\caption{The primary peak, $r_{\rm cc,max}$, of the ICCF/CCF as a function of wavelength, at which the ICCF is measured relative to the 1367~$\angstrom$ light curve in the observed frame. 
Using ICCF method, the symbols are inferred for NGC 5548, while the median and 25\%-75\% quartiles of 200 re-sampled segments with uneven sampling and photometric error are represented by the orange dot-dashed line and the surrounding light-gray region, respectively.
For comparison, the relation implied by, using the CCF method, the median of 200 initial segments with even sampling is presented as the black solid line.
}\label{fig:lc_ccf_rcc}
\end{figure}

\begin{figure*}[tb!]
\centering
\includegraphics[width=\columnwidth]{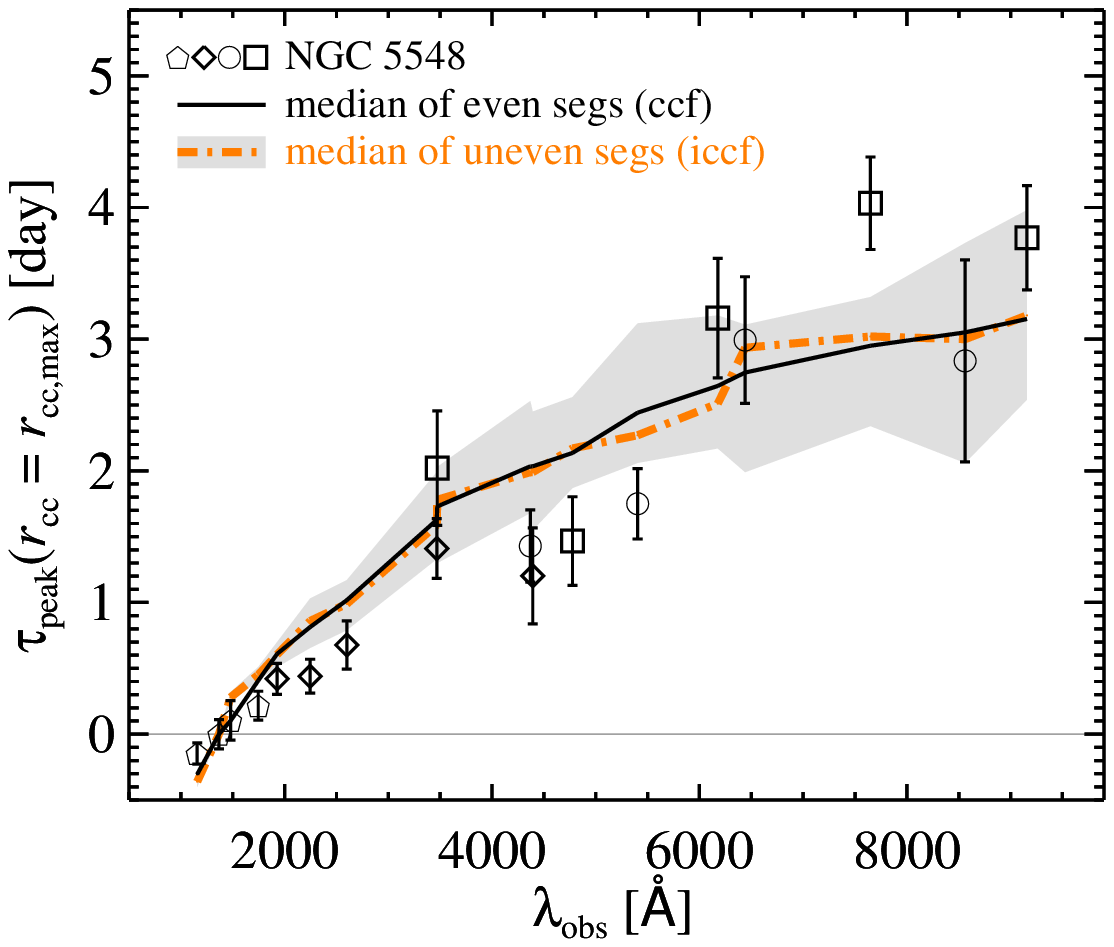}
\includegraphics[width=\columnwidth]{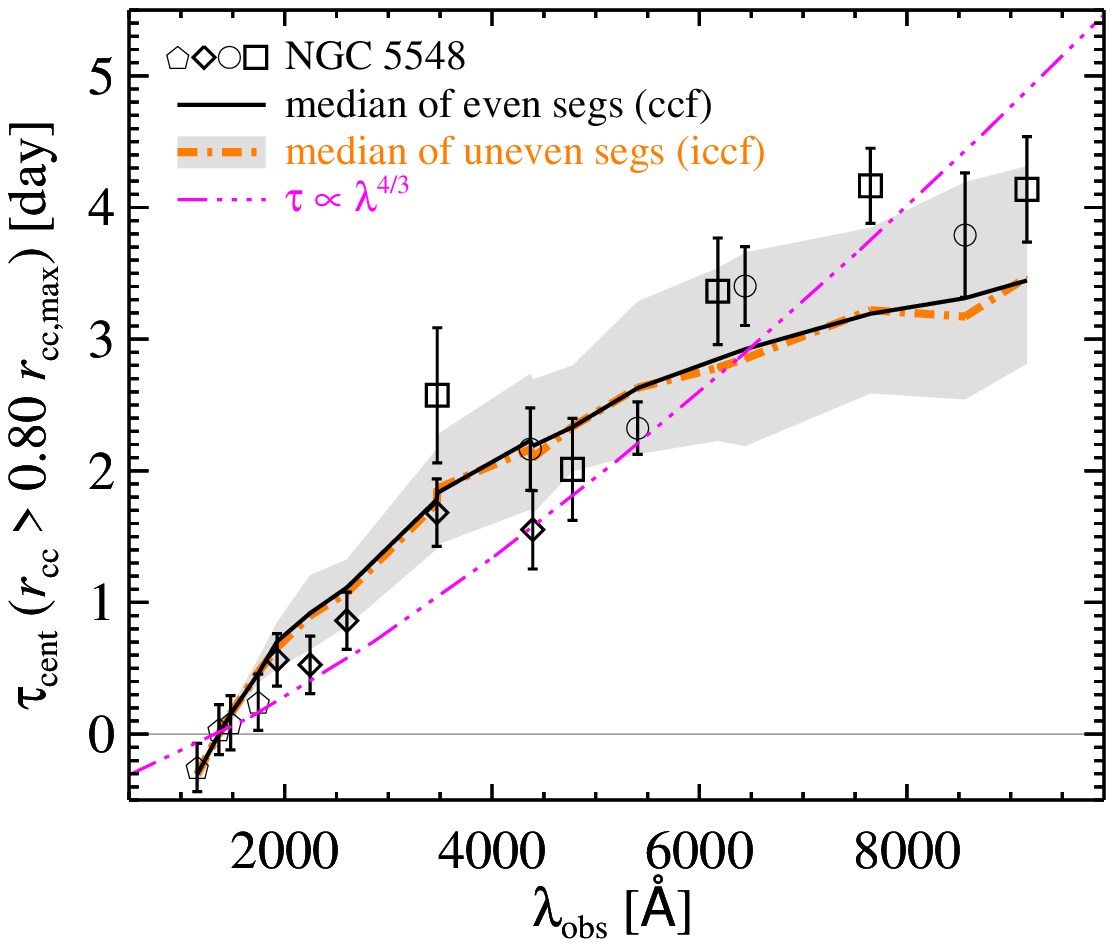}
\caption{The time delay (left panel for the peak lag, $\tau_{\rm peak}$, at the peak of ICCF/CCF and right one for the correlation-weighted mean lag, $\tau_{\rm cent}$, with $r_{\rm cc} > 0.8~\,r_{\rm cc,max}$) as a function of wavelength, at which the light curve is measured relative to the 1367~$\angstrom$ one in the observed frame. The symbols and lines have the same definition as in Figure~\ref{fig:lc_ccf_rcc}.
In the right panel, the lag-wavelength relation, $\tau_{\rm cent} = \tau_0 [(\lambda/1367~\angstrom)^\beta - 1]$, implied by the reprocessing of hard X-rays illuminating a thin disk and fitted by \citet{Fausnaugh2016} with $\tau_0 = 0.42$ days and fixed $\beta = 4/3$, is also added for comparison (the magenta triple-dot-dashed line).
}\label{fig:lc_ccf_tau}
\end{figure*}

These observational properties of NGC 5548 are reported in Figures~\ref{fig:lc_ccf_rcc}-\ref{fig:lc_ccf_wcent} (the same symbols as in Figure~\ref{fig:lc_depen_vs_ngc5548}), while their 1$\sigma$ uncertainties are estimated with $10^3$ realizations using the flux randomization/random subset selection method.
Note that using the same detrending procedure as \citet{Fausnaugh2016}, we can almost reproduce the lag properties for NGC 5548, except the maximal correlation.
Furthermore, we confirm that after applying the same detrending procedure both the peak and centroid lags are mostly reduced by only $\lesssim 0.2$ days for the longest wavelength considered.

On our simulations, the same procedure has been applied to all 200 re-sampled segments with uneven sampling and photometric error. For each segment, the above properties of ICCF as a function of wavelength can be deduced. 
Then, the median and 25\%-75\% quartiles of these properties of the 200 segments are assessed and represented, respectively, by the orange dot-dashed line and the surrounding light-gray region in Figures~\ref{fig:lc_ccf_rcc}-\ref{fig:lc_ccf_wcent}. 
For comparison, also shown as the black solid line is the relation implied, using the CCF method, by the median of these properties of the 200 initial segments with even sampling and without photometric error.

Attractively, a satisfactory agreement between simulation and data for the properties of ICCF/CCF in Figures~\ref{fig:lc_ccf_rcc}-\ref{fig:lc_ccf_wcent} as well as the color variation in Figure~\ref{fig:lc_ccf_theta} could be statistically established, regardless of the simple empirical treatment involved in the current model. Detailed comparisons are delineated in the following.

\subsection{Coordination}\label{sect:coordination}

The nice agreement on maximal correlation, $r_{\rm cc,max}$, between simulation and NGC 5548 is illustrated in Figure~\ref{fig:lc_ccf_rcc}. The correlation strength becomes a little bit smaller once the wavelength difference is larger. The photometric error also works in slightly reducing the correlation.

In \citet{Cai2016} assuming completely independent temperature fluctuation over the whole disk, the implied correlation decreases very quickly with increasing wavelength. However, once a common large-scale temperature fluctuation is speculated over the whole disk, the implied correlation could only gently decrease with increasing wavelength. The strong correlation between bands with large wavelength difference is guaranteed by the common fluctuation, while the gentle decrease is a result of the local independent fluctuations on different disk regions far away in radius.

In current scenario, the most relevant parameters determining the correlation strength 
or the declining behavior with increasing wavelength are $f_{\rm com/indep}$, $\gamma$, and $N_\phi$.
A combination of $f_{\rm com/indep}$ and $\gamma$, i.e., $f_{\rm com/indep} \times (r/r_{\rm in})^\gamma$, represents the relative contribution between common and independent fluctuations as a function of radius.
As shown in the inset of middle panel of Figure~\ref{fig:tempfluc_indep2depen}, the short-term variation ratio of common fluctuations to independent ones, i.e., $\sigma_{\rm s,com}/\sigma_{\rm s,indep} = f_{\rm com/indep} \times (r/r_{\rm in})^{\gamma+\alpha/2}$, is small except at the inner most regions around $\sim 10~r_{\rm g}$, which emits the extreme UV radiation with wavelength of $\lesssim 1000~\angstrom$.
For our concerned wavelength of $\gtrsim 1000~\angstrom$ and disk region of $\gtrsim 30~r_{\rm g}$, a small contribution of common fluctuation is already enough for the strong correlation of light curves over the UV/optical wavelengths. What really works, such that a small contribution of common fluctuation can be persisted and even enhanced in the light curve, is the summing of radiations from several zones near a given radius. The local independent radiations are dimmed by cancellation, while the global common radiations are magnified by addition. Therefore, the number of zones per each layer, $N_\phi$, also adjudges the correlation in such a way that large $N_\phi$ stronger correlation.

By the way, we have not compared the correlation between the hard/soft X-ray and the UV/optical in this paper. This is mainly limited by the fact that the bulk of X-ray emission from corona has not been properly taken into account in the current model, except few implied by the inner most regions of accretion disk. Though having not been modeled, the X-ray emission may be also subject to the common large-scale fluctuation speculated in our model such that there is still some amount of correlation between X-ray and far-UV \citep[e.g., $r_{\rm cc,max} \sim 0.4$ from Table 6 of][]{Fausnaugh2016}, surviving from the quick and dramatic variation of independent X-ray emission. Also noted is that the X-ray fluctuation may not be well-modeled by the CAR(1) process as discussed by \citet{Zhu2016}. Therefore, we prefer a deferred more sophisticated model for dealing with the hard/soft X-ray and their relations to UV/optical.

\subsection{Lag-wavelength relation}\label{sect:lag-wavelength}

Focusing on Figure~\ref{fig:lc_ccf_tau} for the lag-wavelength relation (left panel for $\tau_{\rm peak}-\lambda$ and right one for $\tau_{\rm cent}-\lambda$), the larger lag with increasing wavelength is attractive, comparable to that of NGC 5548.
Some differences can also be found and they may be partly due to the sampling plus measurement errors and/or attributed to the contamination of the continuum emissions from broad line region that have not been properly accounted for within the current model.

The lag implied by our model is a result of the different regression capabilities returning the common large-scale fluctuation to the corresponding mean at different radii with distinctive local damping timescales. In order to explain the observed timescale-dependent color variation, the local damping timescales of temperature fluctuations have been demonstrated to be proportional to the radius, i.e., $\tau \propto r^\alpha$ with $\alpha > 0$ \citep{Cai2016}. Consequently, the common large-scale fluctuations imprinted at smaller radii are more quickly withdrawn than those imprinted at larger radii, and so do the radiations at shorter wavelengths relative to the longer ones. Therefore, there is a positive delay for the emission at a longer wavelength relative to that at a shorter one. 

Another noticeable feature of the implied lag-wavelength relation is the flattening of the lag at long wavelengths. 
This is a manifestation of the limit of regression capability at large radii. With increasing radius, the local damping timescale increases and the regression capability decreases to zero. Since the difference of regression capability at two radii results in the lag of the corresponding emissions, further increasing radius from a large enough radius would only introduce little decrease of the regression capability and therefore little increase of the lag.
The clear flattening of the lag at long wavelengths emerging in our prediction not only seems to be consistent with that hinted by the current data but also is a direct testable prediction of our model.

Prominently, as illustrated in the right panel of Figure~\ref{fig:lc_ccf_tau}, there is a qualitative difference on the increasing trend with wavelength between our scenario (black solid or orange dot-dashed line) and the reprocessing one (magenta triple-dot-dashed line). The latter relation of $\tau_{\rm cent} = \tau_0 [(\lambda/1367~\angstrom)^\beta - 1]$ is implied by the reprocessing of hard X-rays illuminating a thin disk and fitted by \citet{Fausnaugh2016} with $\tau_0 = 0.42$ days and fixed $\beta = 4/3$. 

To highlight this new scenario we have only considered $\alpha = 1$ for the current paper, however, as could be expected, the lag between two given wavelengths would increase with increasing the parameter $\alpha$ of $\tau(r) \sim r^\alpha$.
Moreover, the lag also sensitively depends on the parameter $\tau_{\rm cut}$, smoothing the common fluctuations, because too much common fluctuations at shorter timescales are survived under a smaller $\tau_{\rm cut}$ and therefore smaller lag.
As before, quantitative discussion will be presented in a separate paper (Z. Y. Cai et al. 2018, in preparation).

\begin{figure*}[tb!]
\centering
\includegraphics[width=\columnwidth]{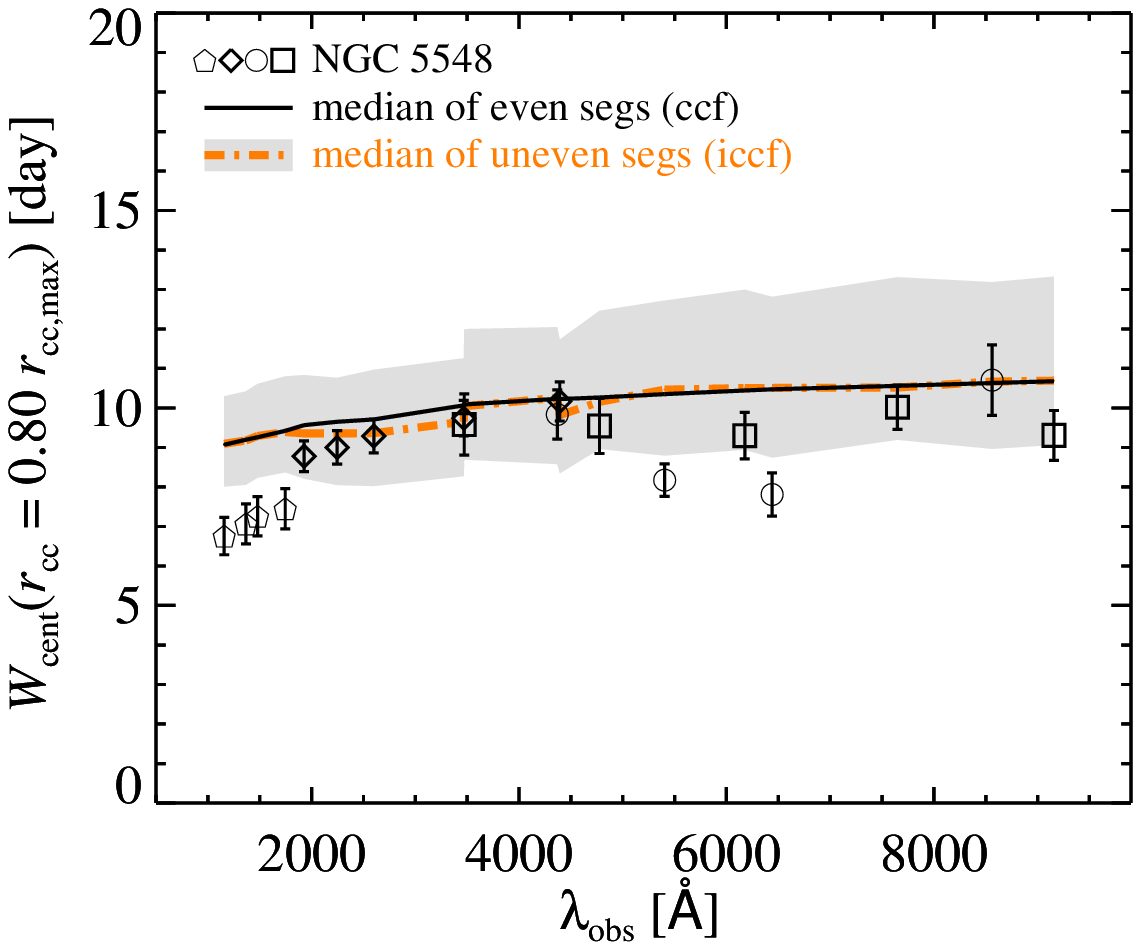}
\includegraphics[width=\columnwidth]{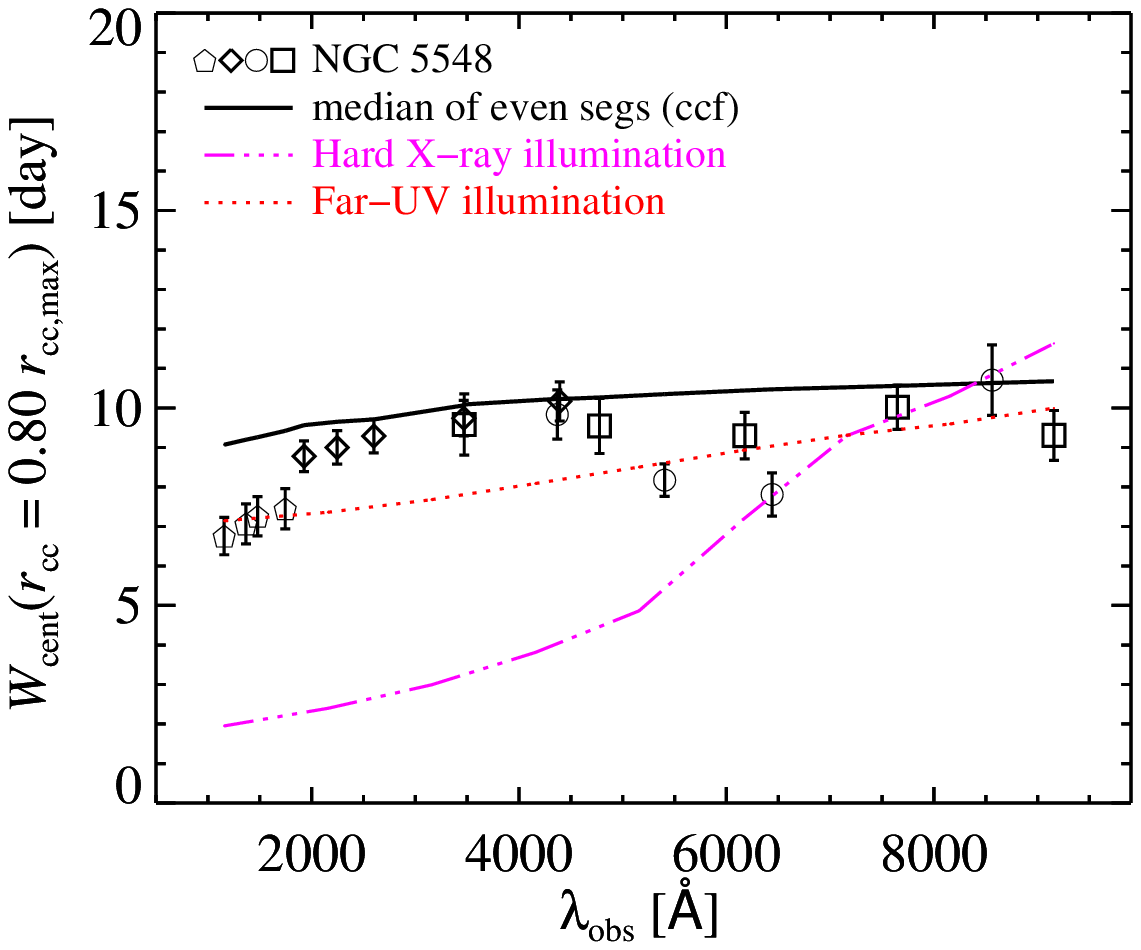}
\caption{The width $W_{\rm cent}$ of the ICCF/CCF at $r_{\rm cc} = 0.8\,r_{\rm cc,max}$ as a function of wavelength, at which the ICCF/CCF is measured relative to the 1367~$\angstrom$ light curve in the observed frame. 
{\it Left panel}: Our models (lines) are compared to that of NGC 5548 (symbols). The symbols and lines have the same definition as in Figure~\ref{fig:lc_ccf_rcc}.
{\it Right panel}: For evenly-sampled light curves with longness of $\sim 180$ days, the width-wavelength relations implied by the hard X-ray reprocessing ({\it Swift} 4.4~$\angstrom$; magenta triple-dot-dashed line) and far-UV reprocessing ({\it HST} 1158~$\angstrom$; red dotted line) are compared to that of our model (black solid line) and that of NGC 5548 (symbols).
}\label{fig:lc_ccf_wcent}
\end{figure*}

\subsection{Correlation width-wavelength relation}\label{sect:width-wave}

Not only the peak of ICCF/CCF of two light curves has important information, but also its width. The latter contains direct hint on the fluctuation quickness of light curves, i.e., quicker fluctuation smaller width.

Here we consider the width $W_{\rm cent}$ of ICCF/CCF at $r_{\rm cc} = 0.8~r_{\rm cc,max}$ as a reference and present the implication of our models compared to that of NGC 5548 in the left panel of Figure~\ref{fig:lc_ccf_wcent}.
A general agreement is found. Note that in the four {\it HST} bands, the fact that the widths of NGC 5548 are slightly smaller than implied by our models just manifests the situation that the {\it HST} light curves of NGC 5548 have more high frequency fluctuations, while our simulations imply relatively smoothed ones (cf. the top portion of Figure~\ref{fig:lc_depen_vs_ngc5548}). 
On one hand, the similar high frequency fluctuations existing on light curves with close wavelengths may be possible if some extra temperature fluctuations happen quickly, especially at smaller radii, and locally (or only being able to propagate over a small region of disk). 
One the other hand, it could indicate that the far-UV variation may not be well-modeled by a simple CAR(1) process as discussed by \citet{Zhu2016}.
Another impressive behavior of the width-wavelength relation is its gentle increase with increasing wavelength. The slightly increasing width with increasing wavelength also indicates the light curves are only slightly more smoothed at longer wavelengths.

To a certain extent, the width is closely related to the lag in our model, that is, smaller width smaller lag. Therefore, it also depends on the three parameters, i.e., $\tau_0$, $\alpha$, and $\tau_{\rm cut}$. The dependence on $\tau_0$ is a trivial time-scaling, while larger $\alpha$ or an increasing $\tau_{\rm cut}$ with increasing radius will result in a quicker increase of width with increasing wavelength. Currently, we have assumed $\alpha = 1.0$ and radius-independent $\tau_{\rm cut}$. Due to their degeneracy, we have checked that it is possible to find an acceptable agreement between data and our model with larger $\alpha$ and smaller $\tau_0$. Moreover, alleviating the radius-independence of $\tau_{\rm cut}$ would also provide more freedom to the model, but its physical mechanism should be clarified in advance.

As already pointed out by \citeauthor{GardnerDone2017} (\citeyear{GardnerDone2017}, see also \citealt{KazanasNayakshin2001,Breedt2009}) that disk reprocessing of the observed hard X-ray light curve produces optical light curves with too much fast variability, the statement is now been well elucidated by the width-wavelength relation, where the width is smaller for much more fast variability. 

To further compare the width-wavelength relations implied by different scenarios, following the procedure introduced in \citet{Zhu2017}, we use the JAVELIN \citep{Zu2011,Zu2013} to interpolate the observed {\it Swift} 4.4~$\angstrom$ hard X-ray and {\it HST} 1158~$\angstrom$ far-UV light curves, assuming them as the driving light curves, and then produces the reprocessed UV/optical light curves. The two observed light curves are both interpolated to a cadence of $\sim 0.01$ days, which is good enough for our current analysis.
Although the observed hard X-ray light curve spans a much longer duration with good sampling ($\sim 870$ days) than that of the {\it HST} one ($\sim 180$ days), we only consider the same duration with an observed time length of $\sim 180$ days as aforementioned in order to compare the widths of ICCF/CCF implied by the same longness of light curve.
Note that to be consistent with the analysis presented in this paper we have only presented the reprocessing results on face-on viewing. Generally, the implied lags by reprocessing scenario are shorter, so we have enlarged the value of $M_\bullet \dot M$ by a factor of 27 such that we can further compare the other properties given the same level of lags. Owing to the potential uncertainties of black hole mass $M_\bullet$ and accretion rate $\dot M$ measurements, this is one of the accessible solutions against the shorter lags implied by the reprocessing scenario \citep{Starkey2017}. Finally, the output UV/optical light curves have further been rescaled, such that their variation amplitudes are the same as the observed ones, before the other properties of light curves are estimated.

The relation implied by disk reprocessing of the observed hard X-ray light curve (the magenta triple-dot-dashed line in the right panel of Figure~\ref{fig:lc_ccf_wcent}) is steeper than that implied by disk reprocessing of the observed far-UV light curve (the red dotted line in the same panel).
The smaller widths at short wavelengths implied by the hard X-ray reprocessing are due to the too short smoothing timescales, i.e., $\lesssim 1-2$ days for light-crossing, to smear out the high frequency variations in the incident hard X-ray light curve.
Motivated by the consistence between the far-UV reprocessing and data, \citet{GardnerDone2017} propose that the fast variability of hard X-rays is suppressed by an intervening structure, or an puffed-up Comptonized disk region, between the inner most hard X-ray emitting region and the outer thin disk. The inner portion of this intervening structure could radiate the low-temperature Comptonized emission and be the origin of the soft X-ray excess, while its outer portion emits the far-UV radiation and can then illuminate the outer thin disk.
Different from their intervening structure to suppress the fast variability, we have in our scenario directly assumed a smoothed common large-scale fluctuation over the disk. However, this smoothed common fluctuation may be a resultant manifestation of the speculated intervening structure (see Section~\ref{sect:phy_mech} for more discussion).

\subsection{Color variation}\label{sect:color_variation}

\begin{figure}[tb!]
\centering
\includegraphics[width=\columnwidth]{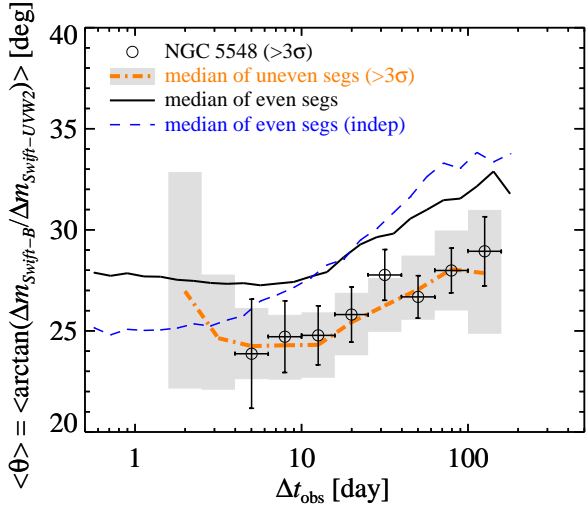}
\caption{The color variation between {\it Swift-UVW2} and {\it Swift-B} bands as a function of the observed time interval.
After having applied the 3 $\sigma$ selection on the variation, the color variation of our model with uneven sampling and photometric error (orange dot-dashed line and surrounding light-gray region for the median and 25\%-75\% quartiles of 200 re-sampled segments, respectively) is compared to that of NGC 5548 (open circles). 
Instead, for the model with even sampling and without photometric error, the median of the color variation of 200 segments with (without) common fluctuation is illustrated as the black solid (blue dashed) line.
}\label{fig:lc_ccf_theta}
\end{figure}

\citet{Sun2014} suggest that the observed (timescale-dependent) color variation is a powerful tool that can be used to effectively rule out several explanations for the quasar UV/optical variation. In \citet{Cai2016}, we have further quantitatively stated that by solely changing the global accretion rate the implied color variation in UV/optical is always timescale-independent, while an inhomogeneous accretion disk model where {\it the damping timescale of temperature fluctuation is radius-dependent} can naturally produce the observed timescale-dependent color variation.
Note that in \citet{Cai2016} the temperature fluctuations in different disk zones are completely independent, while we have introduced in the current model a common large-scale fluctuation superimposed on those independent temperature fluctuations. If the common fluctuation dominates over the independent ones, the disk would return back to the simple case of changing the global accretion rate and therefore the implied color variation may be close to the timescale-independent one.
Since we have discussed in Section~\ref{sect:model} that the common temperature fluctuation only mildly guides the independent ones, the corresponding color variation is expected to be persistently timescale-dependent but also flattened to some extent. To more quantitatively ascertain the color variation implied by the current model and compare with that of NGC 5548, we present in Figure~\ref{fig:lc_ccf_theta} the color variation between {\it Swift-UVW2} and {\it Swift-B} bands for an example.

We adopt the same definition of $\theta$ in magnitude-magnitude space as \citet{Sun2014} for the color variation. Note that smaller $\theta$ more significant bluer-when-brighter. For the real data of NGC 5548 and our simulated light curves with uneven sampling and photometric error, the light curves are not sampled at exactly the same epochs, so we have re-binned the initial light curves in bins of 1 days\footnote{Almost, there is, if any, one initial data point in one bin. Once more data points are found in one bin, the initial fluxes and errors are averaged weighted by the errors.} and only consider the epochs when the light curves are both sampled. Thereafter, we compute $\theta$ in logarithmic bins of width $\Delta \log(\Delta t_{\rm obs}) = 0.2$ and then obtain the mean color variation $\langle \theta \rangle$ in each bin after applying the 3 $\sigma$ criterion on the variation. The open circles in Figure~\ref{fig:lc_ccf_theta} represent the mean color variation of NGC 5548 between the considered two bands, while the corresponding 1$\sigma$ uncertainties are estimated according to the distribution of mean color variations of 10$^3$ realizations obtained using the flux randomization/random subsection selection method. On the other hand, we have rescaled the simulated 200 sources such that their light curves have the same mean and variation in flux density as those of NGC 5548 in the same band, in order to empirically correct for the host and emission line contaminations.
Note that this correction assumes that the host and emission line contaminations have not changed on the concerned timescales and therefore only changes the absolute level of $\theta$ whose timescale-dependent shape is kept. 
We obtain the mean color variation for each of the 200 sources, while the illustrated color variation (orange dot-dashed line) and the uncertainties (light-gray region) in Figure~\ref{fig:lc_ccf_theta}, respectively, represent the median and 25\%-75\% quartiles of the color variations of 200 sources.

Globally, for both data and our simulations after applying the 3 $\sigma$ criterion on the variation, there is an increasing trend with increasing time interval and therefore the color variation is confirmed to be timescale-dependent. Nice agreement between them is intuitive, except in the smallest two bins of time interval.
In those bins, NGC 5548 does not have convincing variations more than 3 $\sigma$, while does only $\sim 70\%$ of the simulated sources (i.e., $\sim 140$ out of 200 in current simulation). Note that this percentage could vary from simulation to simulation, but the behavior of color variation at longer timescales is always preserved. Due to the large photometric uncertainties relative to the inherent small variation on short time interval, the averaged $\theta$ is biased toward 45$^\circ$ \citep{Zhu2016}. The same reason is attributed to the larger averaged $\theta$ if a criterion with fewer $\sigma$ cut has been applied.
This point is again clearly demonstrated in Figure~\ref{fig:lc_ccf_theta} that the color variation, implied by the same 200 simulated sources but with even sampling and without photometric error (i.e., no selection on variation; the solid black line), is systematically offset upward, compared to that implied by the 200 simulated source with uneven sampling, photometric error, and 3 $\sigma$ criterion on the variation (the orange dot-dashed line).

For comparison, we also illustrate the color variation implied by the inhomogeneous accretion disk with completely independent temperature fluctuations (the blue dashed line in Figure~\ref{fig:lc_ccf_theta}). Its steeper slope is also consistent with our expectation discussed previously.
Although the shape of color variation becomes slightly flattened in our current model with a common fluctuation, it is still timescale-dependent. Moreover, the consistence between the current model and data on color variation may instead indicate the need of a common fluctuation, though the explanation with degenerate parameters in model may be possibly responsible for the flattened shape \citep{Cai2016}.

The color variations implied by the hard X-ray/far-UV reprocessing processes have been studied by \citet{Zhu2017} who suggest that the reprocessing models fail to reproduce the observed timescale dependency on the color variation of NGC 5548. Assuming a CAR(1) process to simulate an evenly sampled and long enough driving light curve illuminating the thin disk, the resultant color variation has been elucidated to be timescale-dependent/-independent only below/above a characteristic timescale, which is related to the inter-band time lags. For the black hole mass and Eddington ratio of NGC 5548 together with the two reference bands, i.e., {\it Swift-UVW2} and {\it Swift-B}, this characteristic timescale is $\sim 10$ days, which means that the corresponding color variation implied by the reprocessing process is quite flat at $\Delta t_{\rm obs} \gtrsim 10$ days \citep[cf. the left panel of Figure 3 in][]{Zhu2017}. Instead, our inhomogeneous model implies a continuously increasing color variation beyond $\sim 10$ days.
Prominently, the different tendency of color variations implied by the reprocessing model and our scenario is another testable feature as long as long enough light curve of a single source is available in the up-coming time domain era \citep[e.g.,][]{Ivezic2008,LSST2017}.

\section{Timescales and physical mechanisms}\label{sect:phy_mech}

So far so good, our current scenario, where the assumed common large-scale fluctuation represents a mixing of many distinct outward/inward propagations over the disk, can statistically reproduce many properties of NGC 5548. The characteristic smoothing timescale, $\tau_{\rm cut}$, of the common large-scale fluctuation is related to the average speed of propagations. A smaller smoothing timescale would lead to a smaller lag among UV/optical bands.
Moreover, the inter-band lag implied by our model is mainly attributed to the difference of regression capability related to the local damping timescale of temperature fluctuation, i.e., $\tau(r) \propto r^\alpha$.
In this section, we discuss these timescales and the potential physical mechanisms responsible for our current empirical treatment.

\subsection{Potential origin for the common large-scale fluctuation}

On the accretion disk, there may be many outward/inward propagations happening at different timescales and speeds. A mixing of them could be a net common large-scale fluctuation beyond a characteristic timescale related to an average speed of propagations. Therefore, the smoothing timescale, $\tau_{\rm cut} \simeq 10$ days, of the speculated common large-scale fluctuation is then associated with the average speed of propagations.
For our concerned wavelength range from $\sim 1000~\angstrom$ to $\sim 9000~\angstrom$, the most relevant thin disk ranges from $\sim 30~r_{\rm g}$ to $\sim 550~r_{\rm g}$, given the assumed black hole mass and Eddington ratio for NGC 5548. 
Using the constrained $\tau_{\rm cut}$, the average speed of propagations may be estimated as $\gtrsim \Delta r / \tau_{\rm cut} \simeq 0.15c \times (\Delta r / 500~r_{\rm g})$, where $\Delta r$ is the relevant disk range.
It could be further slightly smaller if the common fluctuation is initialized within the relevant disk range and/or if only a smaller part of the relevant disk range is under the impact of the common fluctuation.
Note that this kind of large propagation speed of fluctuation across the disk may be a solution for the observed changing-look AGNs where a dramatic outburst of $\sim 1-2$ order of magnitude can be generated on a short timescale of $\sim 1-10$ years.

Absolutely, neither the viscous radial drift nor the sound mode on the Shakura-Sunyaev thin disk can set up the common fluctuation. Considering the radiation-pressure-dominated, electron-scattering-dominated inner region\footnote{For standard thin disk  with high-temperature of $T \gtrsim 10^4$ K, the radiation pressure dominates over gas pressure at $r \lesssim 156.4~r_{\rm g} \times  (\alpha/0.1)^{2/21} (M_\bullet/5\times 10^7 M_\odot)^{2/21} (\lambda_{\rm Edd}/0.1)^{16/21} (\eta/0.1)^{-16/21}$ and the electron scattering dominates over free-free absorption at $r \lesssim ~ 5.0 \times 10^3~r_{\rm g} \times (\lambda_{\rm Edd}/0.1)^{2/3} (\eta/0.1)^{-2/3} $, where $r_{\rm g} \equiv GM_\bullet/c^2$ \citep[cf.][]{ShakuraSunyaev1973,KFM2008book}. By simply extending the solution of radiation-pressure-dominated inner region to the gas-pressure-dominated region as we present in Equations~(\ref{eq:t_vis}-\ref{eq:t_s}), the resultant viscous and sound crossing timescales would all be overestimated by about several factors to an order of magnitude, but our discussion is not altered by this crude approximation.}
and simply rescaling the thin disk solution for stellar mass black hole to AGN \citep[][cf. their Equation~3.63]{KFM2008book}, both the viscous timescale
\begin{align}\label{eq:t_vis}
t_{\rm vis} &\equiv \left|\frac{r}{\upsilon_{\rm r}}\right| \simeq (80.5 - 2.1 \times 10^6) \left(\frac{r}{30-550~r_{\rm g}}\right)^{7/2} \nonumber \\
&\left(\frac{M_\bullet}{5\times10^7~M_\odot}\right) \left(\frac{\alpha_{\rm vis}}{0.1}\right)^{-1} \left(\frac{\lambda_{\rm Edd}}{0.1}\right)^{-2} \left(\frac{\eta}{0.1}\right)^{2} f^{-1} ~ {\rm years},
\end{align}
where $\upsilon_{\rm r}$ is the radial drift velocity, $\alpha_{\rm vis}$ is the viscosity parameter, $\eta$ is the radiation efficiency, and $f \equiv 1 - \sqrt{r_{\rm in}/r} \lesssim 1$, and the sound crossing timescale
\begin{align}\label{eq:t_s}
t_{\rm s} \equiv \frac{r}{c_{\rm s}} &\simeq (0.15-217.5) \left( \frac{r}{30-550~r_{\rm g}} \right)^{5/2} \nonumber\\
& \left( \frac{M_\bullet}{5 \times 10^7~M_\odot} \right) \left( \frac{\lambda_{\rm Edd}}{0.1} \right)^{-1} \left( \frac{\eta}{0.1} \right) f^{-1} ~ {\rm years},
\end{align}
where $c_{\rm s}$ is the sound speed, are significantly longer than the observed time delay on order of a few days. 

Circumvention may be found by considering a disk-corona configuration. Although the formation and geometry of the Comptonizing corona is still unclear \citep[e.g.,][]{ReynoldsNowak2003,Liu2009}, its temperature is clearly much larger than that of the underlying thin disk. For example, the temperature of ions in corona could be as large as $\sim 10^{11-12}~{\rm K}$, and the corresponding sound speed $c_{\rm s} \sim 0.1-0.4 c (T/10^{11-12}{\rm K})^{1/2}$ \citep[e.g.,][]{Cox2000book,Netzer2013book}, which is marginally acceptable for the speculated common large-scale fluctuation. 
Although the inward radial drift velocity is quite large for corona, the sound speed can be still significantly larger than the radial drift velocity as long as the viscosity parameter is not too large, e.g., $c_{\rm s}/|\upsilon_{\rm r}| \sim 10 (\alpha_{\rm vis}/0.1)^{-1}$ from a self-similar analysis on hot accretion flow \citep{NarayanYi1994,YuanNarayan2014}.
In this case, the common fluctuation may represent some kind of instabilities happening in the corona and the differential interaction between corona and disk gives rise to the observed time delay.
However, the corona would be required to extend to $\sim 500~r_{\rm g}$ or even larger. This may be in conflict with the extremely compact X-ray size of AGNs, i.e., the half-light radius of $\sim 10~r_{\rm g}$, estimated by microlensing \citep{Chartas2009,Dai2010}. Nevertheless, the corona may be possibly largely extended but with an outer portion too faint to be detected.
Theoretically, given a mass accretion rate of $\sim 0.1~\dot M_{\rm Edd}$ and a normal viscosity parameter for NGC 5548, about $1\%-10\%$ of the accreted mass in thin disk may be evaporated into corona up to radius as large as $\sim 10^3~r_{\rm g}$ \citep[][cf. their Figure~1]{Liu2009}.

Besides the quick propagation of instability in the corona, highly ionized outflows/winds, launching from inner disk regions close to the black hole, may also help transmitting the perturbation by interacting with the accretion disk and then induce a common large-scale fluctuation across the disk. By identifying blueshifted X-ray absorption lines, ultrafast outflows probably located at $\sim 10^{2-4}~r_{\rm g}$ have been found to be quite ubiquitous in AGNs and the median velocity of them is $\sim 0.1c$ with a high-velocity tail up to $\sim 0.3c$ \citep[][and references therein]{Tombesi2010b,Tombesi2013,Gofford2013,KingPounds2015}. 
A mixture of ultrafast outflows generated in the inner disk of $\sim 10^2~r_{\rm g}$ have been speculated to be the origin of disk winds \citep[e.g.,][]{Mou2017}.
Actually, a fast and long-lived outflow has been reported by \citet{Kaastra2014} in NGC 5548, identified instead by broad, blueshifted UV absorption troughs with outflow velocities of $\sim 0.003c$ to $\sim 0.02c$ and plausibly lasting from June 2011 to February 2014. They argue that the broad UV absorption troughs and the X-ray obscuration may arise in the same photoionized gas located at $\sim 10^{16-19}$ cm or $\sim 1.3 \times 10^{3-6}~r_{\rm g}$. Further considering the viewing angle, the velocity of outflows from more inner regions may be as large as those found in X-ray absorption lines.

The other intuitive circumvention may be that the common large-scale fluctuation in our model is a result of the hard X-ray or far-UV reprocessing.
Only considering the illumination onto a static thin disk, the time delay increasing with the UV/optical wavelength is due to the light travel difference, which has a timescale of
\begin{equation}
t_{\rm lc} \simeq (0.1-1.6) \frac{r}{30-550~r_{\rm g}} \frac{M_\bullet}{5\times10^7~M_\odot}~{\rm days},
\end{equation}
and is generally smaller than the observed lags. The extra time delay required to be consistent with the observed one may arise from the reprocessing time, on which the reprocessing process on the disk surface takes place, or/and the relaxation time, on which the disk structure responds to the incident illumination, i.e., from non-equilibrium to equilibrium.
The former is usually neglected \citep[e.g.,][]{Starkey2017}, while the latter has been similarly discussed by \citet{GardnerDone2017}.
In typical reprocessing model, only a small layer of disk surface is heated and promptly re-emitted thermally. If the reprocessing process consists of ionization and recombination, the corresponding timescale as discussed by \citet{Nayakshin2002} may be approximated by, after incorporating the temperature profile $T_{\rm d}$ of disk, 
\begin{align}
	&t_{\rm rep} \simeq (0.8-94.3)  \left( \frac{r}{30-550~r_{\rm g}} \right)^{13/8} \left( \frac{M_\bullet}{5 \times 10^7~M_\odot} \right)^{15/8}  \nonumber\\
	& \left( \frac{\lambda_{\rm Edd}}{0.1} \right)^{1/8} \left( \frac{\eta}{0.1} \right)^{-1/8}  f^{1/8} \frac{\xi}{10^4} Z^{-2} \left( \frac{L_{\rm X}}{10^{46}~{\rm ergs~s^{-1}}} \right)^{-1} ~ {\rm s},
\end{align}
where $\xi$ is the ionization parameter, $Z$ is the charge of ion, and $L_{\rm X}$ is the total X-ray luminosity. 
Therefore, the reprocessing time of X-ray illumination may be safely neglected.

Although the ionization balance can be quickly re-established, the induced perturbation on the vertical disk structure may only be smeared out on a hydrostatic timescale $t_{\rm hyd}$, which is similar to the dynamical timescale $t_{\rm dyn}$, that is \citep[e.g.,][]{Frank2002book}
\begin{align}\label{eq:t_dyn}
	t_{\rm hyd} \simeq t_{\rm dyn} \simeq (0.5-36.8) \left(\frac{r}{30-550~r_{\rm g}}\right)^{1.5} \frac{M_\bullet}{5\times10^7~M_\odot}~{\rm days},
\end{align}
which may be linked to the local damping timescale of temperature fluctuation as discussed in the next subsection. 
In our model, the differential regression of the common fluctuation gives rise to the desired time delay between wavelengths (see Sections~\ref{sect:model}-\ref{sect:results}). Accordingly, the observed lags may be a combination of one induced by the light travel difference and the other one due to the differential regression between different radii.
This is similar to that proposed by \citet{GardnerDone2017}, where the thin disk is heated by the far-UV illumination and then oscillates vertically with transition of disk states. Whether as they suggest the thin disk switches to and back from a Comptonized disk or simply an vertical oscillation of thin disk is interestingly to be explored. 
In this case our smoothing timescale $\tau_{\rm cut}$ may correspond to the smoothing process in the puffed-up Comptonized disk proposed by \citet{GardnerDone2017} to smear out the high-frequency power of the hard X-rays and then produce a relatively smoothed far-UV radiation illuminating the thin disk.
Therefore, the far-UV illumination seems to provide a nice seed for our common fluctuation.

Another interesting point to be mentioned is that the current smaller lags implied by illuminating a static thin disk may be increased to be more consistent with observation by illuminating an inhomogeneous disk (F. F. Zhu et al. 2018, in preparation) because in the latter case the disk size at a given wavelength can be increased by factors of 2-4 \citep{DexterAgol2011}.
Consequently, a postponed quantitatively self-consistent model for illuminating an inhomogeneous disk model with and without the common fluctuation would be worthy of assessing their relative contributions, e.g., the lags, and help comprehending the variabilities across X-ray to UV/optical.

\subsection{Local timescale of temperature fluctuation}

In our model, the lag is mainly attributed to the difference of regression capability related to the local damping timescale of temperature fluctuation, that is,
\begin{equation}
\tau(r) \simeq (5.0 - 91.7) \left(\frac{r}{30 - 550~r_{\rm g}}\right)^\alpha \frac{M_\bullet}{5\times10^7~M_\odot}~{\rm days},
\end{equation}
with $\alpha = 1$. If the local damping timescale is independent of radius, there would be no lag even the common large-scale fluctuation has been considered.

Over the relevant disk range, this timescale is comparable to the dynamical timescale (Equation~\ref{eq:t_dyn}) as well as to the thermal or cooling timescale, that is \citep[e.g.,][]{Siemiginowska1996,CollierPeterson2001,Lawrence2012}
\begin{equation}
t_{\rm th} \simeq (4.9-382.2) \left(\frac{\alpha_{\rm vis}}{0.1}\right)^{-1} \left(\frac{r}{30-550~r_{\rm g}}\right)^{1.5} \frac{M_\bullet}{5\times10^7~M_\odot} ~ {\rm days}.
\end{equation}
Note that regardless the different indexes of radius dependence the dynamical/thermal timescale is marginally smaller/larger than our assumed damping timescale for temperature fluctuation. A combination of these two timescales may be implied by our result. Therefore, the disk may be subject to the dynamical instabilities and/or thermal instabilities.

\section{Conclusions}\label{sect:conclusion}

All in all, serving as a choice to overcome the challenges encountered by the reprocessing model, we have demonstrated in this contribution an interesting phenomenological scenario where the inhomogeneous accretion disk model, whose local {\it independent} temperature fluctuations are subject to a speculated {\it common} large-scale temperature fluctuation, can intrinsically generate the inter-band lag across UV to optical continua, by a nice agreement with several observational properties of NGC 5548, including the tight inter-band correlations and lags as well as the timescale-dependent color variations.
The emergence of lag is the result of {\it a quicker regression capability at smaller radii (i.e., shorter local damping timescale of temperature fluctuation) than at larger ones} when responding to the common large-scale fluctuation\footnote{An interesting analogy may be the popular fashion of dressing style, which is a global property of human kind about aesthetics. There is an easily reachable consensus that the energetic younger people (the local temperature fluctuation with shorter timescale at smaller radius) ordinarily more quickly follows the new style (the global common fluctuation) than the conventional older ones (the local temperature fluctuation with longer timescale at larger radius). Consequently, the older (the emission at longer wavelength from larger radius) lags the younger (the emission at shorter wavelength from smaller radius). }.
This new mechanism would attract more interests if in the future the lag-wavelength relation can be also confirmed for luminous AGNs, where X-ray emission and then the reprocessing process become less predominant.
Over the disk range relevant to the UV/optical continua, the average speed of propagations as large as $\gtrsim 15\%$ of the speed of light may be required by this common fluctuation.
Several potential physical mechanisms for inducing such propagations, including through corona, via outflow, and induced by illumination, are discussed. To better understand the UV/optical continuum variations and the relation between X-ray and UV/optical, a more sophisticated scenario, consisting of the inhomogeneous accretion disk and the reprocessing process, would be deserving to be explored in future work.

\section*{Acknowledgement}

We would like to acknowledge the referee for his/her encouraging comments on this manuscript, his/her time spent to confirm our calculations, and his/her pinpoint intuition bringing to our attention a misleading typo regarding the dependent temperature fluctuation of Equation~(\ref{eq:depen}).
We are grateful to Renbin Yan for the comments on outflow and Heng-Xiao Guo, Da-Bin Lin and Shan-Shan Weng for a careful reading on the manuscript.
This work is partly supported by National Basic Research Program of China (973 program, grant No. 2015CB857005), National Science Foundation of China (grants Nos. 11503024, 11233002, 11421303 $\&$ 11573023), and Specialized Research Fund for the Doctoral Program of Higher Education (20123402110030).
Z.Y.C. acknowledges support from the Fundamental Research Funds for the Central Universities.
J.X.W. thanks support from the Chinese Top-notch Young Talents Program.
M.Y.S. acknowledges support from the China Postdoctoral Science Foundation (grant No. 2016M600485) and the National Science Foundation of China (grant No. 11603022).
Special acknowledgements to all the comments and suggestions from Jason Dexter, Chris Done, Andy Lawrence, Paulina Lira, Nankun Meng, and John J. Ruan, during the conference ``Unveiling the Physics Behind Extreme AGN Variability'' held at the University of the Virgin Islands, St. Thomas Campus from July 10$^{\rm th}$ through July 14$^{\rm th}$ 2017.

\bibliography{ms_qv.bbl}

\end{document}